\documentclass{article}
\usepackage{agd-paper}
\usepackage[letterpaper,margin=1in]{geometry}
\usepackage{authblk}

\newcommand*{\email}[1]{\href{mailto:#1}{#1}}

\begin{document}
\title{The minimal hitting set generation problem: algorithms and computation}

\author{Andrew Gainer-Dewar\thanks{\email{gainerdewar@uchc.edu}}}

\author{Paola Vera-Licona\thanks{\email{veralicona@uchc.edu}}}

\affil{
  Center for Quantitative Medicine,
  UConn Health,
  Farmington, Connecticut, USA
}

\maketitle

\begin{abstract}
  Finding inclusion-minimal \emph{hitting sets} for a given collection of sets is a fundamental combinatorial problem with applications in domains as diverse as Boolean algebra, computational biology, and data mining.
  Much of the algorithmic literature focuses on the problem of \emph{recognizing} the collection of minimal hitting sets; however, in many of the applications, it is more important to \emph{generate} these hitting sets.
  We survey twenty algorithms from across a variety of domains, considering their history, classification, useful features, and computational performance on a variety of synthetic and real-world inputs.
  We also provide a suite of implementations of these algorithms with a ready-to-use, platform-agnostic interface based on Docker containers and the AlgoRun framework, so that interested computational scientists can easily perform similar tests with inputs from their own research areas on their own computers or through a convenient Web interface.
\end{abstract}

\section{Introduction}
\label{s:intro}
Fix a family $S$ of sets $S_{1}, S_{2}, \dots$.
A \emph{hitting set} $T$ of $S$ is a set which intersects each of the sets $S_{i}$; $T$ is a \emph{minimal hitting set} (hereafter \enquote{MHS}) if no proper subset of $T$ is a hitting set of $S$.

The problem of generating the collection of MHSes for a given set family is of interest in a wide variety of domains, and it has been explicitly studied (under a variety of names) in the contexts of combinatorics (\cref{s:probs-combinatorics}, \cite{eiter2008}), Boolean algebra (\cref{s:probs-boolalg}, \cite{hagenthesis,domingo1999,fredman1996}), fault diagnosis (\cref{s:probs-diagnosis}, \cite{reiter1987,cardoso2014,quaritsch2014,wotawa2001,lin2003,abreu2009,pill2012}), data mining (\cref{s:probs-datamining}, \cite{dong2005,bailey2003,vinterbo2000,li2002}), and computational biology (\cref{s:probs-compbio}, \cite{veralicona2013,vazquez2009,hadicke2011,jarrah2007,veralicona2012,ideker2000}), among others.
While some interesting results have been obtained for the associated decision problem, the computational complexity of this problem is currently unknown.
Nevertheless, there is an extensive literature of algorithms to generate minimal hitting sets.

In this paper, we survey the state of the art of algorithms for enumerating MHSes, with particular attention dedicated to publicly-available software implementations and their performance on problems derived from various applied domains.
We begin in \cref{s:cognate} with discussion of the cognate problems that emerge in several applied domains, including explanations of how each can be translated back to our MHS generation problem.
In \cref{s:complexity}, we survey what is known about the complexity of the problem, both in general and in specialized tractable cases.

The bulk of the paper is dedicated in \cref{s:algorithms} to surveying the history of some twenty algorithms for enumerating MHSes.
For each algorithm, we discuss its structure, relevant properties, any known bounds on its complexity, and information about available software implementations.

In \cref{s:experiment}, we present the results of extensive benchmarks run on the available implementations of these algorithms, focusing on running time for large examples derived from specific applications.
Additionally, we provide these implementations in a unified, ready-to-use framework of Docker images based on the AlgoRun framework, which are available to download or to use through a Web interface.
Interested readers and computational scientists may use these containers to try out the algorithms on their own input data and incorporate them into processing pipelines.

We conclude in \cref{s:conclusion} with an overview of our results and the state of the art in MHS generation algorithms.

\section{Cognate problems}
\label{s:cognate}
There are many important problems in numerous domains that can be reduced or translated to MHS problems.
We survey a few of them here, organized by domain.

\subsection{Combinatorics}
\label{s:probs-combinatorics}

\paragraph{Transversal hypergraph problem}
Given a set $V$ of \emph{vertices} and a set $E$ of sets $E_{1}, E_{2}, \dots \subseteq V$ of \emph{hyperedges}, the pair $H = \pair{V}{E}$ is a \emph{hypergraph}.
$H$ is \emph{finite} if $V$, $E$, and all the $E_{i}$ are finite sets.
A hypergraph is \emph{simple} if none of its edges is a subset of any other edge.
The hypergraph $H = \pair{V}{E}$ can naturally be identified with the set family $E$, so the theory of hypergraphs is similar to that of set families.
Readers interested in the full theory of hypergraphs should consult Berge's 1984 monograph \cite{berge1984} on the subject.

In the finite hypergraph setting, a (minimal) hitting set of $E$ is called a \emph{(minimal) transversal} of $H$.
The collection of all transversals of $H$ is its \emph{transversal hypergraph} $\Tr H$.
$\Tr H$ is also sometimes called the \emph{dual} of $H$ because, in the case that $H$ is simple, $\funceval{\Tr}{\Tr H} = H$.
(More generally, $\funceval{\Tr}{\Tr H} = \min H$, the hypergraph obtained from $H$ by removing all non-inclusion-minimal edges.)

There is an extensive literature on the problem of determining whether two hypergraphs $H_{1}$ and $H_{2}$ are transversal of each other.
We will consider a number of algorithms developed for this purpose in \cref{s:algorithms}.
Interested readers can consult the recent survey of Eiter \cite{eiter2008} and Ph.D.~thesis of Hagen \cite{hagenthesis} for more details about this subject.

\paragraph{Set cover problem}
Fix a hypergraph $H = \pair{V}{E}$.
A \emph{set cover} of $H$ is a set $E' \subseteq E$ of edges of $H$ with the property that every vertex in $V$ is in some edge in $E'$.
Of course, $E$ is a set cover, but there may be others, and it is of particular interest to compute the inclusion-minimal ones.

We can construct a new hypergraph $G$ whose vertex set is $E$ and whose edge set is $V$ (that is, with a vertex in $G$ for every edge of $H$ and an edge in $G$ for each collection of edges in $H$ with a common vertex).
Then the MHSes of $G$ are exactly the minimal set covers of $H$.
Thus, an algorithm for generating MHSes can be used to enumerate minimal set covers, and vice versa.

\paragraph{Independent set problem}
Fix a hypergraph $H = \pair{V}{E}$.
An \emph{independent set} of $H$ is a set $V'$ of vertices of $H$ with the property that no edge of $H$ is a subset of $V'$.
Of course, the empty vertex set $\emptyset$ is a set cover, but there may be others, and it is of particular interest to compute the inclusion-maximimal ones.
It is easy to show that an independent set is the complement of a hitting set and thus that a maximal independent set is the complement of a minimal hitting set.
Thus, we can enumerate maximium independent sets of $H$ simply by applying some algorithm for MHS generation and then taking complements.

\subsection{Boolean algebra}
\label{s:probs-boolalg}

\begin{definition}
  \label{def:boolfunc}
  A \emph{Boolean function} is a function $f = f(x_{1}, x_{2}, \dots, x_{k}): \bool^{k} \to \bool$, where $\bool = \set{0, 1}$ and $k$ is a non-negative integer.
\end{definition}

A $k$-ary Boolean function is described by an algebraic expression, called a \emph{Boolean expression} or \emph{Boolean formula}, which consists of the binary variables $x_{1}, \dots, x_{k}$, the binary \emph{conjunction operator} $\land$ (often written \tand), the binary \emph{disjunction operator} $\lor$ (often written \tor), and the unary \emph{negation operator} $\lnot$ (often written \tnot).

If a Boolean function $f$ has the property that $f(X) \leq f(Y)$ (resp.~$f(X) \geq f(Y)$) for any inputs $X \leq Y$, we say that $f$ is \emph{positive} (resp.~\emph{negative}).
A function which is either positive or negative is \emph{monotone}.
Monotone Boolean functions appear in a wide variety of formal and applied settings.
(See \cite{korshunov2003} for an extensive survey.)

We can easily ensure that this property holds by restricting the formulas considered:

\begin{definition}
  \label{def:boolfunc-monotone}
  A formula for a Boolean function is \emph{monotone} if it contains only variables, conjunctions, and disjunctions (i.e.~no negations).
\end{definition}

\begin{theorem}
  \label{thm:boolfunc-monotone}
  Any monotone formula gives a monotone Boolean function.
\end{theorem}

However, representation in these propositional formulas is not unique; the same function may be given by many formulas, even after commutativity is considered.
Accordingly, restrictions are often placed on the formulas to ensure uniqueness.

\begin{definition}
  \label{def:boolfunc-normal}
  A formula for a Boolean function is in \emph{conjunctive normal form} (CNF) if it is a conjunction of disjunctions of literals and \emph{disjunctive normal form} (DNF) if it is a disjunction of conjunctions of literals.
\end{definition}

\begin{example}
  \label{ex:boolfuncex-normal}
  The formula $\pbrac*{x_{1} \lor x_{2}} \land x_{3}$ is in CNF, while $\pbrac*{x_{2} \land x_{3}} \lor \pbrac*{x_{1} \land x_{3}}$ is in DNF.
\end{example}

Both normal forms are of great interest for computational applications because they are easy to decompose and compare.
In addition, a given monotone Boolean function has exactly one each of CNF and DNF formulas satisfying an \emph{irredundancy} condition.

If two given formulas $C$ in CNF and $D$ in DNF represent the same function, we say they are \emph{dual}.
Deciding whether this is true of a given pair $(C, D)$ is widely studied in the literature of complexity theory; see the excellent survey in the Ph.D.~thesis of Hagen \cite{hagenthesis}, which gives this problem the picturesque name \probname{monet}.
We may also consider the generation problem \probname{mongen}, which asks for the DNF formula $D$ equivalent to a given CNF formula $C$.

Surprisingly, \probname{mongen} is computationally equivalent to the MHS generation problem.
To illustrate why, consider the function $\pbrac*{x_{2} \lor x_{3}} \land \pbrac*{x_{1} \lor x_{3}}$, which is evidently in CNF.
Its \emph{clauses} (i.e.~the disjunctive components) are formed from the variable sets $\set{x_{2}, x_{3}}$ and $\set{x_{1}, x_{3}}$.
Since these clauses are conjoined, to satisfy the formula we must satisfy each clause, which means we must set to $1$ at least one variable in each clause; for example, we might take $x_{2}$ and $x_{1}$ or take $x_{3}$ alone.

Indeed, this is exactly equivalent to finding hitting sets of the set family $\set{\set{2, 3}, \set{1, 3}}$.
Moreover, the irredundancy requirement for a DNF is equivalent to requiring the hitting sets to be minimal.
The collection $\set{\set{2, 1}, \set{3}}$ of MHSes corresponds to the DNF $\pbrac*{x_{2} \land x_{1}} \lor x_{3}$ of the function.

\subsection{Model-based fault diagnosis}
\label{s:probs-diagnosis}
Modern engineered systems may involve incredible numbers of interconnected components; for example, the recently-retired NASA Space Shuttle reportedly had over 2.5 million moving parts.
When such a system fails to perform as intended, it is infeasible to check or replace every part and connection; thus, \emph{diagnostic} procedures are needed to narrow attention to some subset of components which may have caused the observed failure.
In a celebrated 1987 paper \cite{reiter1987}, Reiter developed the foundation for a formal theory of model-based diagnosis (MBD), which we will introduce briefly.

Consider a system made up of some finite set $V$ of components, each of which may be either \emph{active} or \emph{inactive} during any given \emph{transaction} or activity.
We make a series of \emph{observations} of transactions of the system, recording which components are active and whether the behavior is normal or anomalous.
If any of the observed transactions are anomalous, we assume this is due to one or more faulty components.
A \emph{diagnosis} of the faulty system is a set $D$ of components which, if all are faulty, would explain all the anomalous transactions.
Under a parsimony restriction, the interesting diagnoses are those which are inclusion-minimal or irredundant.
(Reiter only applies the term \enquote{diagnosis} to these minimal examples.)

Let $F$ be the set of faulty transactions, where each such transaction is represented as the set of components involved.
A diagnosis is then a hitting set of $F$, and the parsimonious diagnoses are exactly the MHSes of $F$.
Thus, any algorithm for generating MHSes is directly applicable to fault diagnosis.
Reiter proposed an algorithm for exactly this purpose; we will discuss it and its successors in \cref{s:algorithms}.

\subsection{Computational biology}
\label{s:probs-compbio}

Minimal hitting sets have appeared as an important combinatorial motif in numerous problems in computational biology.
We survey four of them here.

\paragraph{Minimal cut sets in metabolic networks}
A \emph{metabolic reaction network} is the system by which metabolic and physical processes that determine the physiological and biochemical properties of a cell, are represented. As such, these networks comprise the chemical reactions of metabolism, the metabolic pathways and the regulatory interactions that guide these reactions. In its graph representation, a metabolic network is a directed hypergraph in which each of $m$ vertices represents a \emph{metabolite} and each of $n$ directed hyperedges represents a \emph{biochemical reaction}. Information about the reactions can be encoded in a \emph{stoichiometric matrix} $M$ in which each row represents a metabolite, each column represents a biochemical reaction, and the entry $S_{i, j}$ represents the coefficient of metabolite $i$ in reaction $j$.
If the coefficient $S_{i, j}$ is positive, metabolite $i$ is produced by reaction $j$; if the coefficient is negative, the metabolite is consumed; and if the coefficient is zero, the metabolite is not involved in the reaction at all and thus is not in its hyperedge.
An $m$-dimensional vector may be employed to represent concentrations of metabolites, while an $n$-dimensional vector may represent rates of reactions.

It is typically assumed that some \emph{internal} metabolites are at steady state and thus must have a net zero rate of change in the system.
Such a steady state corresponds to an $n$-dimensional \emph{flux vector} $\vec{x}$ with the property that $S \vec{x} = \vec{0}$; we call a vector satisfying this condition an \emph{admissible flux mode}.
We call such a flux mode \emph{elementary} if its \emph{support} (the set of reactions with nonzero fluxes) is inclusion-minimal.

The notion of elementary flux modes (\enquote{EFMs}) was introduced by Schuster and Hilgetag in \cite{schuster1994}; subsequent work has developed numerous techniques from linear algebra and computational geometry to find the EFMs.
 see the recent surveys \cite{trinh2009,zanghellini2013} for overview of the problem, the methods and software which are used to solve it, and various applications.

One important area of applications of metabolic network analysis is \emph{metabolic engineering}, in which the metabolic network of an organism is modified to adjust its production.
In \cite{klamt2004}, Klamt and Gilles focus on blocking a target reaction through \emph{cut sets}, which they define as a set of reactions whose removal from the network leaves no feasible balanced flux distribution involving the target reaction.
To do this, they first compute all the elementary flux modes which involve the target reaction.
They then construct a set family whose elements are the reactions of the original network and whose sets are the relevant elementary flux modes.
Finally, they compute the minimal hitting sets of this family.
Subsequently, Haus \textit{et al.} developed in \cite{haus2008} a specialized version of the \algname{FK-A} algorithm (cf.~\cref{s:alg-fk}) which greatly improves on other methods available at the time.

\paragraph{Optimal combinations of interventions in signal transduction networks}
Signal transduction describes the process of conversion of external signals to a specific internal cellular response, such as gene expression, cell division, or even cell suicide. This process begins at the cell membrane where an external stimulus initiates a cascade of enzymatic reactions inside the cell that typically includes phosphorylation of proteins as mediators of downstream processes. A \emph{signal transduction network} is represented as a signed directed graph in which each of the vertices is a \emph{signaling component} (such as a protein, gene in a cell) and each edge represents an interaction which the source induces in the target (either positive-signed \emph{activation} or negative-signed \emph{inhibition}).
Biological signaling networks typically \cite{zevedei2005} exhibit a natural decomposition into \emph{input}, \emph{intermediate}, and \emph{output} nodes; engineering and control of these networks then typically depends on adjusting the input and intermediate layers to obtain some outcome at the outputs.
As an analogous of elementary flux modes (\enquote{EFMs}) in metabolic networks, in \cite{wang2011}, Wang and Albert introduced the notion of \emph{elementary signaling modes} (ESMs). ESMs are minimal sets of signaling components which can perform signal transduction from inputs to outputs; ESMs are the natural analogues of EFMs for signal transduction networks.
They also provide algorithms for generating the ESMs of a given network.

Once the topology and ESMs of a signal transduction network are known, it may be of interest to control how signal flows from a given set of \emph{source} nodes to a given set of \emph{targets}, perhaps avoiding interfering with certain \emph{side effect} nodes along the way.
Vera-Licona \textit{et al.} introduced the OCSANA framework to study this problem in \cite{veralicona2013}.
They begin by computing the ESMs which pass from the specified sources to the specified targets.
They then construct a set family whose elements are the signalling components and whose sets are these ESMs.
They next compute MHSes of this family, which they term \emph{combinations of interventions} (CIs).
Finally, they apply their \enquote{OCSANA} scoring, which encodes heuristics about control of the targets and influence on the side effects, and use the results to identify the most promising CIs.

Since discovery of hitting sets is a crucial step in the algorithm, the authors of \cite{veralicona2013} performed an experimental comparison.
In particular, they tested Berge's algorithm (cf.~\cref{s:alg-berge}) and an approach similar to \algname{MTMiner} (cf.~\cref{s:alg-hbc}) which incorporates the OCSANA score into the generation process.
They found their new algorithm to improve substantially on Berge's algorithm.

\paragraph{Reverse-engineering biological networks from high-throughput data}
In some situations, the network structure itself is not known.
In this case, it is useful to reconstruct the network from experimental data through \emph{reverse engineering}.
Broadly, the \emph{biological reverse-engineering problem} is that of \enquote{analyzing a given system in order to identify, from biological data, the components of the system and their relationships} \cite{dasgupta2010}.
This may result in either a topological representation of the network structure (typically expressed as an enriched graph) or a full dynamic model (in terms of a mathematical  model).
Numerous approaches to the topological reverse-engineering problem have been introduced, including at least two which use MHS generation techniques.
See \cite{dasgupta2010} for a comparative survey of these two approaches and the relative performance of the specialized algorithms developed for each; we give here only a brief overview of each.

\subparagraph{Ideker}
In 2000, Ideker \textit{et al.} introduce a method to infer the topology of a network of gene regulatory interactions in \cite{ideker2000}.
They first perform a series of experiments in which various genes in the system are forced to \emph{high} or \emph{low} expression and use standard microarray techniques to measure the expression of the other genes.
For each gene $i$, they then consider all pairs $\set{a, b}$ of experiments for which that gene's expression values differ, then find the set $S_{a, b}$ of all other genes whose expression values also changed.

Under their assumptions, one or more genes in $S_{a, b}$ must cause the change in $i$ between experiments $a$ and $b$.
A set of genes which intersects \emph{all} the sets $S_{a, b}$ is a candidate to explain the observed variation in $i$ over the whole suite of experiments, and thus to be the set of genes connected to $i$ in the regulatory network.
Thus, they generate a collection of hitting sets for the family $S_{a, b}$; in particular, they develop an algorithm based on the standard \emph{branch and bound} optimization technique which gives sets of minimal cardinality.
Since this collection may be large, they use an entropy-based approach to iteratively generate new experiments which will discriminate among the candidates and re-apply the algorithm to the enlarged data set to improve the accuracy of the predicted networks.

\subparagraph{Jarrah}
In 2007, Jarrah \textit{et al.} introduce another method to infer the topology of a gene regulatory network in \cite{jarrah2007} which focuses on time series data within a single experiment.
They associate to each gene $i$ a variable $x_{i}$ from some finite field $k$, then represent each time point $t$ in the experiment as an assignment $\mathbf{x}^{t}$.
For a given gene $i$, they then consider how the values of $\mathbf{x}^{t}$ determine the value $x_{i}^{t+1}$ for each $t$.

Under their assumptions, if $x_{i}$ is observed to take different values at times $t_{1}$ and $t_{2}$, it must be due to some other variables being different at times $t_{1} - 1$ and $t_{2} - 1$.
Thus, for each pair $t_{1} \neq t_{2}$ with $x_{i}^{t_{1}} \neq x_{i}^{t_{2}}$, they construct a set $S_{t_{1}, t_{2}} = \setbuilder{j}{j \neq i, x_{j}^{t_{1} - 1} \neq x_{j}^{t_{2} - 1}}$ which encodes the variables which may be responsible for the observed change in $x_{i}$.
Like Ideker et al., they then compute minimal hitting sets of this collection, but their algorithm is formalized in terms of computational algebra and gives a complete enumeration of the family of MHSes.
(We discuss this approach in \cref{s:alg-primdecomp}.)
They also present a heuristic scoring function which may help to select the most viable model from the generated hitting sets.

\paragraph{Drug cocktail development}
Many widely-used antibiotics are effective against some bacterial strains but ineffective against others.
Thus, in cases where more than one strain may be present or the specific strain is unknown, it is necessary to deploy a \enquote{cocktail} of several drugs to increase the number of strains covered.
A similar situation applies in cancer chemotherapy, where different chemotherapeutic agents are known to be effective only against certain cell lines.
In either case, it is desirable to minimize the number of drugs used at once, to minimize the cost of the therapy and the risk of emergent multiple-drug-resistant strains.

Given a set of drugs (say, antibiotics) and a set of targets (say, bacterial strains), we can assign to each target the set of drugs that are effective against it.
A (minimal) hitting set of this set family is then a (minimal) cocktail of drugs that, taken together, affects all the targets.

This application has been studied in detail by Vazquez in \cite{vazquez2009}, using a greedy algorithm to search for very small effective combinations from the NCI60 collection (\cite{shoemaker2006}) of 45334 drugs and 60 cancer cell lines.
He is then able to recommend some of these small MHSes as targets for further research.

\subsection{Data mining}
\label{s:probs-datamining}

A number of problems in data mining can be formalized in terms of MHS generation.
We survey two of them here.

\paragraph{(In)frequent itemset discovery}
One fundamental problem in data mining, introduced by Agrawal \textit{et al.} in \cite{agrawal1993} and developed further in \cite{agrawal1996}, is the discovery of \emph{frequent itemsets} in a database of transactions.
We adopt the formal setting of the problem presented by Boros \textit{et al.} in \cite{boros2003}.
Fix a finite set $A$ of $m$ \emph{transactions}, each of which is a finite subset of a set $I$ of $n$ \emph{items}.
Fix an integer \emph{threshold} $1 \leq t \leq m$.
A set $C$ of items is \emph{$t$-frequent} if at least $t$ transactions are supersets of $C$ and \emph{$t$-infrequent} if no more than $t$ transactions are supersets of $C$.
The \emph{maximal frequent} (resp.~\emph{minimal infrequent}) \emph{itemset problem} is to enumerate inclusion-maximal (resp.~inclusion-minimal) sets $C$ which are $t$-frequent (resp.~$t$-infrequent).

Let $F_{t}$ denote the hypergraph whose edges are maximal $t$-frequent itemsets in $A$ and $I_{t}$ denote the hypergraph whose edges are minimal infrequent itemsets in $A$.
(Both have the common vertex set $I$.)
Any element of $I_{t}$ must intersect the complement of every element of $F_{t}$, and in fact as hypergraphs we have that $I_{t} = \Tr \pbrac{\comp{F_{t}}}$ exactly.
Thus, if either $I_{t}$ or $F_{t}$ is known, the other can be computed using any algorithm for MHS generation.
This connection is explored by Manilla and Toivonen in \cite{mannila1996}; more algorithmic details are given by Toivonen in \cite{toivonen1996}.
An application of these ideas to database privacy is given by Stavropoulos \textit{et al.} in \cite{stavropoulos2015}.

Furthermore, so-called \enquote{joint-generation} algorithms inspired by the \algname{FK} algorithms of Fredman and Khachiyan \cite{fredman1996} (cf.~\cref{s:alg-fk}) can generate $I_{t}$ and $F_{t}$ simultaneously.
This is developed by Gunopulos \textit{et al.} in \cite{gunopulos1997} and its complexity implications explored by Boros \textit{et al.} in \cite{boros2002itemset}.

\paragraph{Emerging pattern discovery}
Another important data mining problem is the discovery of \emph{emerging patterns}, which represent the differences between two subsets of the transactions in a database.
We adopt the formal setting of the problem introduced by Bailey \textit{et al.} in \cite{bailey2003}.
Consider two sets $A$ and $B$ of \emph{transactions}, where each transaction is itself a set of \emph{items}.
A \emph{minimal contrast} is an inclusion-minimal set of items which appears in some transaction in $A$ but no transaction in $B$.
Fix a transaction $a \in A$ and construct a set family whose underlying elements are the items in $a$ and with a set $a \setminus b$ for each $b \in B$.
Then the minimal contrasts supported by $a$ are exactly the minimal hitting sets of this set family.
Thus, any algorithm for MHS generation can be applied to emerging pattern discovery.
Indeed, two of the algorithms we study, \algname{DL} (\cref{s:alg-dl}, \cite{dong2005}) and \algname{BMR} (\cref{s:alg-bmr}, \cite{bailey2003}), were developed for this purpose.

\subsection{Minimal Sudoku puzzles}
\label{s:sudoku}
The \emph{Sudoku} family of puzzles is widely published in newspapers and magazines and is played by millions worldwide.
An instance of Sudoku is a $9 \times 9$ grid of boxes, a few of which already contain digits (\enquote{clues}) from the range 1--9; a solution is an assignment of digits to the remaining boxes so that each of the nine $3 \times 3$ subgrids and each row and each column of the whole puzzle contains each digit exactly once.
An example with seventeen clues is shown in \cref{fig:sudoku}.

\begin{figure}[htbp]
  \centering
  \setlength{\sudokusize}{5cm}
  \renewcommand*\sudokuformat[1]{\Large\sffamily#1}
  \begin{sudoku}
    | | | |8| |1| | | |.
    | | | | | | | |4|3|.
    |5| | | | | | | | |.
    | | | | |7| |8| | |.
    | | | | | | |1| | |.
    | |2| | |3| | | | |.
    |6| | | | | | |7|5|.
    | | |3|4| | | | | |.
    | | | |2| | |6| | |.
  \end{sudoku}
  \caption{An example Sudoku puzzle with 17 clues}
  \label{fig:sudoku}
\end{figure}

Of course, not every possible placement of clues into the grid yields a valid puzzle.
There may be inherent contradictions, such as two identical clues in the same column, so that the puzzle has no solutions.
There may also be ambiguities, in which more than one solution is possible.
A mathematical question of particular interest, then, is: \emph{what is the smallest number of clues in an unambiguous valid Sudoku puzzle?}
Many such puzzles with 17 clues are known, but none with 16 have ever been identified.
In \cite{mcguire2014}, McGuire \textit{et al.} show that an exhaustive search for such puzzles can be formulated in our terms by constructing set families that represent the effects of clues in solved puzzles and then searching for hitting sets of size $16$ or less.
They ran this search on a supercomputing cluster and proved conclusively that there are no $16$-clue Sudoku puzzles.
They use an algorithm similar to \algname{HST} from \cite{wotawa2001}, discussed in detail in \cref{s:alg-hst}; for speed, they modify the algorithm to essentially build the set families and their hitting sets simultaneously.

\section{Complexity results}
\label{s:complexity}

\subsection{Asymptotic complexity}
\label{s:asymptotic}
Before considering specific algorithms for MHS generation, we should consider the current state of knowledge about the asymptotic complexity of the problem.
It has been known since Karp's seminal 1972 paper \cite{karp1972} that the problem of determining whether a given set family has a hitting set of size no greater than some $k$ is NP-complete.
However, we are more concerned with the collection of all hitting sets than with the existence of a single one.
We therefore consider two other separate but related problems.

\subsubsection{Recognition problem}
\label{s:comp-rec}
Much of the literature on complexity analysis focuses on \enquote{decision problems}, which must have a \enquote{yes} or \enquote{no} answer.
The natural decision variant of the MHS problem is \emph{recognition}: given two hypergraphs $H$ and $G$, to decide whether $H = \Tr G$.
Fredman and Khachiyan present in \cite{fredman1996} an algorithm (discussed in \cref{s:alg-fk}) which tests this in time $n^{\compo{\log n}}$ (for $n$ the sum of the number of hyperedges in $F$ and $G$).
This time bound is notable in that it is \emph{quasi-polynomial}---it is worse than a polynomial bound, but better than an exponential bound or even certain sub-exponential bounds.
The \algname{BM} algorithm introduced by Boros and Makino in \cite{boros2009} improves on this bound in parallel cases.
It is a long-standing open problem to determine whether recognition is possible in polynomial time.

\subsubsection{Generation problem}
\label{s:comp-gen}
For many applications, however, we need to generate the MHSes of a given set family rather than recognize them.
It is straightforward to show that no algorithm can do this in time polynomial in the size of the input.
Consider the example of the \emph{matching graph} $M_{n} = \set{\pair{1}{2}, \pair{3}{4}, \dots, \pair{2n-1}{2n}}$ as a set family.
A minimal hitting set contains a choice of one of the two elements of each edge; thus, there are evidently $2^{n}$ of them.
Simply writing out this result therefore requires at least $\compo{2^{n}}$ time, so no algorithm can  in subexponential time in general.
This is not necessarily to say that the MHS generation problem is intractable; rather, it suggests that it is inappropriate to analyze its complexity input-polynomiality solely in terms of input size.

Johnson \textit{et al.} introduced a formalism to deal with this issue in \cite{johnson1988}.
Instead of considering only the size of the \emph{input} to the problem, we can incorporate the size of the \emph{output} as well.
If a given set family $S$ has $n$ sets and $m$ MHSes, an algorithm for generating those MHSes is said to be \emph{output-polynomial} (or to run in \emph{polynomial total time}) if its running time is $\compo{\funceval{\poly}{n+m}}$.
Unfortunately, this is not known to be achieved by any current algorithm, and Hagen showed in \cite{hagen2009lower} that several important algorithms are \emph{not} output-polynomial.

If we shift our attention to incremental generation, we may instead ask whether an algorithm can generate one MHS at a time with reasonable delay between outputs.
Johnson \textit{et al.} introduced two suitable formalisms in \cite{johnson1988}.
First, a generation algorithm may run in \emph{incremental-polynomial time}; in this case, given a set family and a set of MHSes for it, it should yield a new MHS in time polynomial in the combined size of both of these inputs.
Second, the algorithm may run with \emph{polynomial delay}; this stronger variant of incremental-polynomial time requires that the time required depend only on the size of the set family and \emph{not} on the number of MHSes already known.
Crucially, if an algorithm runs with polynomial delay, it is guaranteed to run in output-polynomial total time, but incremental-polynomial time gives no such guarantee (\cite{johnson1988}).

Incremental time analysis is of particular interest in some classes of applications, where we may wish to generate only some MHSes or to perform further processing on each one as it emerges.
No algorithm is known to solve the MHS generation problem in incremental polynomial time (much less with polynomial delay) in general, but there are some interesting special cases, considered in \cref{s:fpt}.

\subsection{Tractable cases}
\label{s:tractable}
Crucially, the complexity results in \cref{s:asymptotic} concern the performance of algorithms in general, which is to say, for the class of all possible hypergraphs or set families.
Another thread of research has focused on demonstrating that, in such restricted cases, much better complexity results are possible.

\subsubsection{Fixed-parameter tractability}
\label{s:fpt}
In some cases, algorithms are available which \enquote{factor out} some of the complexity of the problem with respect to a particular parameter of the hypergraphs.
Specifically, letting $k$ be the parameter of interest and $n$ be the number of edges of the hypergraph, we may find an algorithm that generates all MHSes in time $\funceval{f}{k} \cdot n^{\compO{1}}$ for some arbitrary function $f$ (which is to say, the time is polynomial in $n$ once $k$ is fixed, though it may depend arbitrarily on $k$).
In this case, we say that the problem is \emph{fixed-parameter tractable} (\enquote{FPT}) with respect to that parameter $k$, since fixing $k$ yields a complexity function that depends polynomially on $n$.

Fixed-parameter tractability results have been obtained for the transversal hypergraph recognition problem with a wide variety of parameters, including
vertex degree parameters (\cite{khachiyan2007computing,domingo1999,mishra1997,elbassioni2008fpt,mishra1997}),
hyperedge size or number parameters (\cite{eiter2003,elbassioni2008fpt,khachiyan2007global}), and
hyperedge intersection or union size parameters (\cite{khachiyan2005,elbassioni2008fpt}).
For a more complete survey, the interested reader may consult \cite[\S 4, \S 7]{hagenthesis}.

\subsubsection{Acyclicity}
\label{s:fpt-acyclic}
A graph is \emph{acyclic} if it contains no cycles---that is, if no non-self-repeating path in the edges leads back to where it starts.
Beeri \textit{et al.} introduced in \cite{beeri1983} a notion of acyclicity in hypergraphs, now known as \emph{$\alpha$-acyclicity}, in the context of the study of relational database schemes.
Fagin subsequently introduced in \cite{fagin1983} the notions of \emph{$\beta$-acyclicity} and \emph{$\gamma$-acyclicity}, which are successively more restrictive and correspond to desirable tractability problems in databases.
Eiter showed that the transversal recognition is solvable in polynomial time for $\beta$-acyclic hypergraphs in \cite{eiter1995} and for $\alpha$-acyclic hypergraphs in \cite{eiter2003}.

\subsection{Limited nondeterminism}
\label{s:np}
Another important line of inquiry for studies of complexity is how its solution improves with \emph{nondeterminism}---that is, if the algorithm is allowed access to some \enquote{free} information.
The crucial question is how many nondeterministic bits are required to achieve a better solution.
Kavvadias and Stavropoulos showed in \cite{kavvadias2003} that the recognition problem is in the class $\compclass{co-NP} \sbrac{\log^{2} n}$ for $n$ the total number of edges in $H$ and $\Tr H$, meaning that only $\compO{\log^{2} n}$ nondeterministic bits are required to demonstrate that two hypergraphs are not transversals of each other.
Since $\log^{2} n$ is subpolynomial, this suggests that the recognition problem is not as hard as the well-known $\compclass{NP}$-complete problems.

\section{Existing algorithms}
\label{s:algorithms}
A wide array of algorithms have been developed to generate MHSes (either explicitly or in the language of various cognate problems).
If we strip away the details of the various domains and applications by casting all the algorithms in the language of MHS generation, we find that they fall naturally into a few high-level categories:
\begin{description}
\item[set iteration appproaches]
  which work through the input set family one set at a time, building MHSes as they go;

\item[divide and conquer approaches]
  which partition the input family into disjoint subfamilies, find their MHSes separately, and then combine them;

\item[MHS buildup approaches]
  which build candidate MHSes one element at a time, keeping track of un-hit sets as they go; and

\item[full cover approaches]
  which improve on the divide-and-conquer approach with a technical hypergraph lemma that allows more efficient recombination.
\end{description}

We survey these categories, including discussions of a few published algorithms for each.
A summary of these algorithms, giving their taxonomic classifications, original problem domains, and relevant characteristics is presented in \cref{tab:alg-feature}.
Whenever possible, we use the terminology of set families and minimal hitting sets, since this language is typically the most straightforward to understand.

It can be shown (cf.~\cite{bioch1995}) that an algorithm for the transversal hypergraph recognition problem (cf.~\cref{s:probs-combinatorics}) can be used to generate MHSes of a given set family with a number of runs that is polynomial in the size of the transversal hypergraph.
This conversion is possible because, if the input hypergraphs $H$ and $G$ are \emph{not} transversals of each other, any recognition algorithm must return a \enquote{witness} of this, which can be translated into an edge which must be added to either $H$ or $G$.
Thus, beginning with some given set family $S$ and an empty collection $T$ of MHSes, we can interpret $S$ as a hypergraph and apply any recognition algorithm to find a new MHS to add to $T$, then repeat until eventually the complete collection is generated and the algorithm confirms $S = \Tr T$.
As a result, we will consider algorithms for both recognition and generation interchangeably.

For nearly all of these algorithms, software implementations are available to perform the calculations on a computer.
We have collected eighteen of these implementations into a public repository at \url{github.com/VeraLiconaResearchGroup/MHSGenerationAlgorithms}.
Source code and information about the implementations are available there.
In addition, we provide a ready-to-use Docker container for each using the AlgoRun framework (cf.~\cite{AlgoRun}) and a Web interface to instances of the software running on the AlgoRun project's servers at \url{algorun.org}.
This framework is used for experimental benchmarks which are presented in \cref{s:experiment}.

\subsection{Set iteration approaches}
\label{s:alg-set}
One type of approach to computing hitting sets of a set family $S$ is to begin with a small subfamily of $S' \subsetneq S$, find the MHSes for $S'$, and then iteratively add more sets to $S'$ and update the MHS collection.
The methods in this section all follow this approach; they differ in how they select the subfamilies $S'$ and in the details of how they update the MHS collection.

\subsubsection[Berge]{Berge (1984)}
\label{s:alg-berge}
The first systematic algorithm for computing transversals of hypergraphs was presented by Berge in \cite{berge1984}, a monograph on the theory of hypergraphs.
Although this algorithm is called the \enquote{Sequential Algorithm} in some literature, we will refer to it as \algname{Berge}.
The core idea of the algorithm is proceed inductively over the hyperedges of the hypergraph, alternately adding a new edge to the intermediate hypergraph under consideration and extending the known transversals.

We first introduce three important operations on hypergraphs.
\begin{definition}
  \label{def:hyp-join}
  Let $H_{1} = \pair{V_{1}}{E_{1}}$ and $H_{2} = \pair{V_{2}}{E_{2}}$ be two hypergraphs.
  Their \emph{vee} $H_{1} \vee H_{2}$ is the hypergraph with vertex set $V = V_{1} \cup V_{2}$ and edge set $E = E_{1} \cup E_{2}$.
  Their \emph{wedge} $H_{1} \wedge H_{2}$ is the hypergraph with vertex set $V = V_{1} \cup V_{2}$ and edge set $E = \setbuilder{e_{1} \cup e_{2}}{e_{1} \in E_{1}, e_{2} \in E_{2}}$.
\end{definition}

\begin{definition}
  \label{def:hyp-min}
  Let $H$ be a hypergraph with vertex set $V$ and edge set $E$.
  Its \emph{minimization} (or \emph{simplification}), $\min H$, is the hypergraph with vertices $V$ and edges $\setbuilder{e \in E}{\nexists f \in E \setminus e: e \subset f}$.
  In other words, $\min H$ retains exactly the inclusion-minimal edges of $H$.
  $H$ is \emph{simple} if $H = \min H$.
\end{definition}

These two operations interact nicely with the transversal construction:

\begin{lemma}
  \label{lem:trans-join}
  Let $H_{1}$ and $H_{2}$ be hypergraphs.
  Then the following relations hold:
  \begin{equation}
    \label{eq:trans-vee}
    \Tr \pbrac{H_{1} \vee H_{2}} = \min \pbrac*{\Tr H_{1} \wedge \Tr H_{2}}
  \end{equation}
  and
  \begin{equation}
    \label{eq:trans-wedge}
    \Tr \pbrac{H_{1} \wedge H_{2}} = \min \pbrac*{\Tr H_{1} \vee \Tr H_{2}}.
  \end{equation}
\end{lemma}

Algorithm \algname{Berge} then proceeds as follows:
\begin{enumerate}
\item
  Let $H$ be a hypergraph with edge set $E = \set{e_{1}, e_{2}, \dots, e_{n}}$ (for an arbitrarily chosen ordering), and for each $i$ let $H_{i}$ be the subhypergraph of $H$ with all its vertices and its first $i$ edges $e_{1}, \dots, e_{i}$.

\item
  For each $i$ in order, compute $\Tr H_{i}$ inductively: $\Tr H_{1} = \setbuilder{v}{v \in e_{1}}$, and $\Tr H_{i} = \min \pbrac*{\Tr H_{i-1} \wedge \Tr e_{i}} = \min \setbuilder{t \cup \set{v}}{t \in \Tr H_{i-1}, v \in e_{i}}$ by \cref{eq:trans-vee}.

\item
  When the iteration is finished, $\Tr H_{n} = \Tr H$ by construction.
\end{enumerate}

Pseudocode for the algorithm (in the lanugage of set families) is given in \cref{code:berge}.

\begin{algorithm}
  \caption{Berge's algorithm\label{code:berge}}
  \begin{algorithmic}[1]
    \Require{A finite set family $S = \set{s_{1}, s_{2}, \dots, s_{n}}$}
    \Ensure{The set of MHSes of $S$}
    \Statex
    \Function{Berge}{$S$}
    \Let{$T$}{$\setbuilder{\set{e}}{e \in s_{1}}$}
    \ForAll{$s \in S \setminus s_{1}$}
    \Let{$T$}{$\setbuilder{t \cup \cbrac{e}}{t \in T, e \in s}$} \label{code:berge:l:build}
    \Let{$T$}{$\min T$} \label{code:berge:l:min} \Comment{Remove non-minimal elements of $T$}
    \EndFor
    \State \Return{$T$}
    \EndFunction
  \end{algorithmic}
\end{algorithm}

As suggested in \cite{hadicke2011}, Berge's algorithm can be adapted to search only for MHSes of cardinality bounded by some $k$ by simply discarding candidates larger than $k$ at \cref{code:berge:l:build,code:berge:l:min} in each stage of the algorithm.

This algorithm is straightforward to implement in code and to study theoretically.
Unfortunately, it also has the potentially to be extremely inefficient.
The complexity is studied thoroughly by Boros \textit{et al.} in \cite{boros2010}.
In particular, for a set family $S$ with $n$ sets and a total of $m$ underlying elements, if the sets are ordered so that the collection $T$ reaches maximum size $k$ during the algorithm, the running time of this algorithm is $\compO{k m n \cdot \funceval{\min}{m, k}}$.

Accordingly, it is clear that the ordering of the edges matters a great deal, since this determines the value of $k$.
Takata showed in \cite{takata2007} that there exists a family of hypergraphs for which \emph{no} edge ordering yields output-polynomial running time, and thus that \algname{Berge} is not output-polynomial in general, even if the edge ordering is optimal.
Boros \textit{et al.} demonstrate in \cite{boros2010}, however, that the worst case is still sub-exponential.

We provide a C++ implementation of \algname{Berge} which supports enumeration of small MHSes in the repository.

\subsubsection[HS-DAG]{Reiter (1987) and Greiner \textit{et al.} (1989)}
\label{s:alg-hsdag}
As discussed in \cref{s:probs-diagnosis}, Reiter introduced the formal theory of model-based diagnosis as an application of MHS enumeration in \cite{reiter1987}.
His approach proceeds through a set family inductively, alternately picking a set which is not yet hit and an element which hits it, until a hitting set for the whole family is found.
It then backtracks to the most recent step where another valid choice was available and repeats.
The intermediate data are stored in a structure Reiter calls a \enquote{hitting set tree}.

However, it was shown by Greiner \textit{et al.} in \cite{greiner1989} that this algorithm is incomplete; the hitting sets it generates are guaranteed to be minimal, but in certain circumstances some MHSes may be missed.
They repair the algorithm, but in the process they sacrifice the acyclicity of Reiter's hitting set tree; the result is a \emph{directed acyclic graph}.
We will refer to the algorithm as \algname{HS-DAG} (for Hitting Set Directed Acyclic Graph).
It is straightforward to implement and is widely studied and cited in the MBD literature.

The authors are not aware of a formal complexity analysis of this algorithm.

It is possible to search only for hitting sets of bounded cardinality with \algname{HS-DAG} simply by restricting the depth of the search DAG.

A Python implementation of this algorithm by the authors of \cite{quaritsch2014} is available in the repository.

\subsubsection[HST]{Wotawa (2001)}
\label{s:alg-hst}
Wotawa returned to Reiter's approach in \cite{wotawa2001}, reviewing the \algname{HS-DAG} algorithm of \cite{greiner1989} (see \cref{s:alg-hsdag}) and adjusting it to reduce the number of set containment checks required.
These improvements render the DAG generalization unnecessary, so the underlying data structure is once again a hitting set tree as originally envisioned by Reiter.
We will refer to the algorithm as \algname{HST} (for Hitting Set Tree).

The authors are not aware of a formal complexity analysis of this algorithm.

It is possible to search only for hitting sets of bounded cardinality with \algname{HST} simply by restricting the depth of the search tree.

A Python implementation of this algorithm by the authors of \cite{quaritsch2014} is available in the repository.

\subsubsection[DL]{Dong and Li (2005)}
\label{s:alg-dl}
Dong and Li considered in \cite{dong2005} the \enquote{emerging patterns problem} discussed in \cref{s:probs-datamining}.
Although their work was essentially independent of the literature on hypergraph transversals, they developed an algorithm thematically very similar to \algname{Berge}.
Their algorithm incorporates some optimizations to the minimization calculation in \cref{eq:trans-wedge} to speed up the loop step.
We will refer to the algorithm as \algname{DL} (for its authors).

However, the running time of \algname{Berge} is dominated by the need to search the intermediate transversals, not the complexity of generating them, so the \algname{DL} optimization should not be expected to improve the worst-case behavior of \algname{Berge}.
Hagen shows in \cite{hagenthesis,hagen2009lower} that Takata's time bounds on \algname{Berge} in \cite{takata2007} apply to \algname{DL} as well, so it is not output-polynomial.
Nevertheless, for families with relatively few sets, \algname{DL} performs well, so it is useful as a subroutine to be used in base cases of other algorithms.

A C implementation of this algorithm by the authors of \cite{murakami2014} is available in the repository.

\subsubsection[BMR]{Bailey \textit{et al.} (2003)}
\label{s:alg-bmr}
Continuing with the study of emerging patterns, Bailey et al. developed in \cite{bailey2003} an algorithm which decomposes the input set family more carefully than Berge's algorithm.
Rather than simply considering one new set at a time, their approach attempts to partition the set family into components with few sets, then use the \algname{DL} algorithm of \cite{dong2005} as a subroutine to compute their MHSes before combining them using \cref{eq:trans-vee}.
We will refer to their algorithm as \algname{BMR} (for Bailey, Manoukian, and Ramamohanarao).

Hagen shows in \cite{hagenthesis,hagen2009lower} that \algname{BMR} is not output-polynomial.
Furthermore, he shows that its complexity is $n^{\compomega{\log \log n}}$, where the $\Omega$ indicates that this is a lower bound instead of the upper bound indicated by $O$ and where $n = \abs{H} + \abs{\Tr H}$.

A C implementation of this algorithm by the authors of \cite{murakami2014} is available in the repository.

\subsubsection[KS]{Kavvadias and Stavropoulos (2005)}
\label{s:alg-ks}
Returning to the explicit study of hypergraph transversals, Kavvadias and Stavropoulos introduced in \cite{kavvadias2005} another algorithm, which seeks to reduce the memory requirements of \algname{Berge} with two optimizations.
First, they preprocess the input set family to combine elements which occur only in the same sets.
Second, they carefully reorganize the processing steps so that many intermediate MHSes can be forgotten without jeopardizing the correctness of the algorithm, allowing them to output MHSes early in the algorithm's run and then discard them.
We will refer to this algorithm as \algname{KS} (for its authors).

The algorithm is designed to run in polynomial memory by avoiding regeneration of candidate hitting sets.
Hagen shows in \cite{hagenthesis,hagen2009lower} that \algname{KS} does not run in output-polynomial time.
Furthermore, he shows that its complexity is $n^{\compomega{\log \log n}}$, where $n$ is the sum of the number of sets in the family and the total number of minimal hitting sets that it admits.

The organization of the search routine in \algname{KS} makes it possible to search for hitting sets of bounded cardinality to save time.
The authors are not aware of an implementation that offers this feature.

A Pascal implementation of this algorithm by the authors of \cite{kavvadias2005} is available in the repository.

\subsection{Divide and conquer approaches}
\label{s:alg-div-conq}
Another type of approach to computing MHSes of a set family $S$ is to partition $S$ into several subfamilies, find their MHSes separately (perhaps recursing until the subfamilies are sufficiently small), and then combine the results.
The algorithms in this section all follow this approach; they differ primarily in how they partition $S$.

\subsubsection[BOOL]{Lin and Jiang (2003)}
\label{s:alg-bool}
Lin and Jiang return in \cite{lin2003} to the problem of model-based diagnosis.
They cast the problem in the Boolean algebra framework, but their algorithm is a straightforward example of the divide-and-conquer approach.
We will refer to this algorithm as \algname{BOOL} (since Lin and Jiang call it the \enquote{Boolean algorithm}).
Their recursive decomposition algorithm proceeds as follows.

\begin{enumerate}
\item
  Let $S$ be a finite set family.
  If $\abs{S} < 2$, it is trivial to find the MHSes of $S$ directly.
  Thus, we assume that $\abs{S} \geq 2$.

\item
  If there is an element $e$ which is present in every set $s \in S$, construct a new set family $S' = \setbuilder{s \setminus e}{s \in S}$.
  Recursively find the MHSes of $S'$, add $\set{e}$, and return the result.

\item
  If there is a set $s \in S$ with $\abs{s} = 1$, let $e$ be the unique element of $s$ and construct a new set family $S' = S \setminus s$.
  Recursively find the MHSes of $S'$, add $e$ to each, and return the result.

\item
  Otherwise, choose some $e \in \bigcup S$ arbitrarily.
  Let $S_{1} = \setbuilder{s \setminus e}{s \in S, e \in s}$ and $S_{2} = \setbuilder{s}{s \in S, e \notin s}$.
  Recursively find the MHSes of $S_{1}$ and $S_{2}$.
  Add $e$ to each MHS of $S_{2}$, then take the union of the results and return.
\end{enumerate}

Pseudocode for the algorithm is given in \cref{code:bool}.
They call this the \enquote{Boolean algorithm}; we will denote it hereafter by \algname{BOOL}.

The Boolean algorithm was subsequently optimized by Pill and Quaritsch in \cite{pill2012}.
In particular, improved its performance in cases that only MHSes of size bounded by some $k$ are desired.

\begin{algorithm}
  \caption{The Boolean algorithm (\algname{BOOL})\label{code:bool}}
  \begin{algorithmic}[1]
    \Require{A finite set family $S = \set{s_{1}, s_{2}, \dots, s_{n}}$}
    \Ensure{The set of MHSes of $S$}
    \Statex
    \Function{Bool}{$S$}
    \Let{$E$}{$\bigcup_{s \in S} s$}
    \If{$\abs{S} = 0$}
    \Let{$T$}{$\emptyset$}
    \ElsIf{$\abs{S} = 1$}
    \State let $S = \set{s}$
    \Let{$T$}{$s$}
    \ElsIf{there is some $e \in E$ such that $e \in s \forall s \in S$}
    \Let{$T$}{$\set{e} \vee \Call{Bool}{\setbuilder{s \setminus e}{s \in S}}$}
    \ElsIf{there is some $s \in S$ such that $\abs{s} = 1$}
    \Let{$G$}{$s \wedge \Call{Bool}{S \setminus s}$}
    \Else
    \State choose $e \in E$ \Comment{can be arbitrary}
    \Let{$S_{1}$}{$\setbuilder{s \setminus e}{s \in S, e \in s}$}
    \Let{$T_{1}$}{$\Call{Bool}{S_{1}}$}
    \Let{$S_{2}$}{$\setbuilder{s}{s \in S, e \notin s}$}
    \Let{$T_{2}$}{$\set{e} \wedge \Call{Bool}{S_{2}}$}
    \Let{$T$}{$T_{1} \cup T_{2}$}
    \EndIf
    \State \Return{$T$}
    \EndFunction
  \end{algorithmic}
\end{algorithm}

The authors are not aware of a formal complexity analysis of this algorithm.

If desired, this algorithm can search for hitting sets of bounded cardinality to save time.

A Python implementation of this algorithm by the authors of \cite{quaritsch2014} is available in the repository.

\subsubsection[FK]{Fredman and Khachiyan (1996)}
\label{s:alg-fk}
Fredman and Khachiyan introduced two iterative algorithms in \cite{fredman1996} to study the recognition version of the MHS problem in the Boolean algebra context.
Like \algname{BOOL}, these two algorithms both proceed by choosing one element, considering sets that do and do not contain that element separately with recursive calls, and then combining the results.
However, they first apply several algebraically-motivated degeneracy tests.
If the tests fail, a new hitting set can be found very efficiently.
If, however, they succeed, it guarantees that an element can be found which is present in many (specifically, logarithmically many) sets but missing from many others.
Considering the sets which do and do not contain this element separately decomposes the problem into two large disjoint sub-problems, which can be considered recursively; the large size of each subproblem ensures that the recursion does not go too deep.
This bound on the recursion depth allows Fredman and Khachian to prove running-time bounds which are the strongest known on any sequential algorithm to date.
We will refer to these two algorithms as \algname{FK-A} and \algname{FK-B} (\enquote{FK} for the authors, who use the names \algname{A} and \algname{B} in \cite{fredman1996}).

The first algorithm, \algname{FK-A}, runs in time $n^{\compO{\log^{2} n}}$ and is relatively straightforward to implement.
The algorithm is modified in \cite{khachiyan2006} to improve its runtime slightly and adapt it to MHS generation.
An implementation in (compiled) C is provided by those authors and is available in the repository.

The second algorithm, \algname{FK-B}, runs in time $n^{\compO{\log n}}$.
(More exactly, its time bound is $n^{4 \funceval{\chi}{n} + \compO{1}}$ where $\funceval{\chi}{n}^{\funceval{\chi}{n}} = n$.)
However, its implementation is significantly more complex than that of \algname{FK-A}.
As a result, most authors (including \cite{khachiyan2006}) have disregarded \algname{FK-B} in comparative studies.
However, analysis in \cite{hagen2009fk} suggests that this assumption may be inaccurate.
The authors are aware of no publicly-available implementations of \algname{FK-B}.

If desired, both algorithms can search for hitting sets of bounded cardinality to save time.
The authors are not aware of an implementation that supports this feature.

The algorithms also allow for \enquote{joint generation} of a set family and its MHSes if subsets of both are known.
This can be advantageous in situations where the set family is not known \textit{a priori} but it is possible to check whether a given set is a member of the family.
For example, Haus \textit{et al.} apply this approach in \cite{haus2008}, as discussed in \cref{s:probs-compbio}.

\subsubsection[STACCATO]{Abreu and Gemund (2009)}
\label{s:alg-staccato}
Model-based diagnosis often involves finding hitting sets of extremely large set families, so approximation algorithms are particularly attractive in this field.
Abreu and Gemund presented such an algorithm in \cite{abreu2009}.
They use a divide-and-conquer approach similar to that of \algname{BOOL}, but which considers the elements in an order determined by a statistical heuristic.
They also define mechanisms to stop the algorithm early to obtain a partial set of approximately minimal hitting sets.
We will refer to this algorithm as \algname{STACCATO} (the name used by its authors in \cite{abreu2009}).

The authors of \cite{abreu2009} claim that, for a set family with $N$ sets and $M$ total elements, the algorithm guarantees to find a hitting set of cardinality $C$ in $\compO{\pbrac*{M \cdot \pbrac*{N + \log M}}^{C}}$ worst-case time and $\compO{C \cdot M}$ space, with improved expected times based on their heuristic and tested experimentally.

If desired, this algorithm can search for hitting sets of bounded cardinality to save time.

A Python implementation of this algorithm by the authors of \cite{quaritsch2014} is available in the repository.

\subsubsection[ParTran]{Leiserson \textit{et al.} (2010)}
\label{s:alg-partran}
Some recent research has focused on parallelizing the search for minimal hitting sets.
Leiserson \textit{et al.} cast this issue in a very abstract setting in \cite{leiserson2010}, developing a framework to parallelize any algorithm that searches for minimal elements of a poset and then applying it to the lattice of hitting sets of a set family.
We will refer to this algorithm as \algname{ParTran} (the name used by its authors in \cite{leiserson2010}).

Treated as a sequential algorithm by running it in a single execution thread, \algname{ParTran} is similar to \algname{BOOL}; its primary distinction is that the two subfamilies $S_{1}$ and $S_{2}$ are carefully chosen to be of similar size to improve parallel efficiency.
The authors are not aware of a formal analysis of its complexity in either sequential or parallel settings.

A Cilk++ implementation of this algorithm by the authors of \cite{leiserson2010} is available in the repository.

\subsubsection[Knuth]{Knuth (2011)}
\label{s:alg-knuth}
\emph{Binary decision diagrams} (BDDs) are a graph-based structure for representing boolean functions and hypergraphs originally introduced by Bryant in \cite{bryant1986}.
Given a set family $S$, it is computationally expensive to \emph{compress} $S$ into a BDD or to \emph{decompress} that BDD back into $S$.
Nevertheless, BDDs are a powerful data structure for certain combinatorial algorithms.
Many logical operations on hypergraphs, such as the $\wedge$ and $\vee$ operations of \cref{def:hyp-join}, are inexpensive to perform on their BDDs.
In exercises 236 and 237 of \cite[\S 7.1.4]{knuth2011}, Knuth asks the reader to devise an algorithm for MHS generation using BDD operations, and in the solutions he presents a simple one.
We will refer to this algorithm as \algname{Knuth}.

The authors are not aware of a formal complexity analysis of \algname{KNUTH}, and Knuth asserts that the worst-case runtime is unknown.

A C implementation of this algorithm by the author of \cite{toda2013} is available in the repository.

\subsubsection[HTC-BDD]{Toda (2013)}
\label{s:alg-htcbdd}
In 2013, Toda improved on the \algname{KNUTH} algorithm in \cite{toda2013} by incorporating a variation on the BDD data structure--the \emph{zero-suppressed binary decision diagram} (ZDD).
After compressing a given set family $S$ into a ZDD, Toda recursively applies a simple divide-and-conquer algorithm to obtain a BDD of all hitting sets of $S$.
He then uses a minimization algorithm to obtain a ZDD of the MHSes of $S$, which he finally decompresses.
We will refer to this algorithm as \algname{HTC-BDD} (the name given to it by Toda).

Toda gives a formal complexity analysis of \algname{HTC-BDD} in \cite{toda2013}, but the resulting bounds are expressed in terms of the intermediate BDD and ZDD data structures and are incommensurable with bounds like those known for \algname{FK-A} and \algname{FK-B}.
One important factor is that the decompression of the output from ZDD format into a list of sets can be very time-consuming.
Details are explored in \cref{s:results}.
However, the ZDD intermediate data structure makes it possible to determine the number of MHSes \emph{without} decompressing, which may be of interest for some applications.

A C implementation of this algorithm by the author of \cite{toda2013} is available in the repository.

\subsubsection[MHS2]{Cardoso and Abreu (2014)}
\label{s:alg-mhs2}
Cardoso and Abreu revisted the \algname{STACCATO} approach in \cite{cardoso2014}.
They present several optimizations to reduce wasted computation.
In addition, their new algorithm is distributed using the widely-used Map-Reduce paradigm, so in principle it can be deployed over very large message-passing distributed computing systems.
It is also designed so that early termination will return a useful approximate result; a collection of hitting sets will be obtained, although they may not be minimal and some may be missing.
We will refer to this algorithm as \algname{\MHStwo} (the name given to it by its authors).

The authors are not aware of a formal analysis of the complexity of \algname{\MHStwo}.

A C++ implementation of the algorithm by the authors of \cite{cardoso2014} is available in the repository.

\subsection{MHS buildup approaches}
\label{s:alg-buildup}
A third type of approach to computing MHSes of a set family $S$ is to construct sets of elements which are expected or guaranteed to be subsets of MHSes, then iteratively add elements until they are hitting sets.
This approach fits into the standard scheme of \enquote{backtracking} combinatorial algorithms.
The approaches in this section all follow this approach; they differ primarily in the conditions used to identify candidate sub-MHSes and the strategies used to avoid redundant calculation.

\subsubsection[MTMiner]{H\'{e}bert \textit{et al.} (2007)}
\label{s:alg-hbc}
H\'{e}bert \textit{et al.} take an approach in \cite{hebert2007} that brings insights from data mining to bear on the MHS generation problem.
We follow the explanation of the algorithm in \cite{elbassioni2014}, which avoids the algebraic complexity of the original.
We will refer to this algorithm as \algname{MTMiner} (the name given to its software implementation by its authors); it is also called \algname{HBC} (for H\'{e}bert, Bretto, and Cr\'{e}milleux) in some literature (e.g.~\cite{elbassioni2014}).

Fix a set family $S = \set{s_{1}, s_{2}, \dots, s_{n}}$ with underlying element set $E = \bigcup S = \set{e_{1}, e_{2}, \dots, e_{m}}$.
The \algname{MTMiner} algorithm is initialized with the set $C_{1} = \setbuilder{\set{e}}{e \in E}$ of element sets of size $1$.
At each step of the algorithm, the set $C_{i}$ of candidate hitting sets of size $i$ is processed.
First, any set in $C_{i}$ which is a hitting set is removed and outputted; as will be seen, it is guaranteed to be minimal.
The remaining sets in $C_{i}$ are extended by combining all pairs $\pair{a}{b}$ which overlap in $i-1$ elements into their union $a \cup b$.
For each of these extended sets (of size $i+1$), the algorithm checks whether more sets are hit by $a \cup b$ than by $a$ or $b$.
If so, $a \cup b$ is added to $C_{i+1}$.
The algorithm terminates no later than $i = n$, by which time all MHSes have been output.

Pseudocode of the algorithm is given in \cref{code:hbc}.

\begin{algorithm}
  \caption{MTMiner algorithm\label{code:hbc}}
  \begin{algorithmic}[1]
    \Require{A family of sets $S = \set{s_{1}, s_{2}, \dots, s_{n}}$}
    \Ensure{The set of MHSes of $S$}
    \Statex
    \Function{MTMiner}{$S$}
    \Let{$C_{1}$}{$\emptyset$} \Comment{initial candidate set}
    \ForAll{$e \in \bigcup S$}
    \If{$e \in s$ for all $s \in S$}
    \State \Output{$\set{e}$}
    \Else
    \Let{$C_{1}$}{$C_{1} \cup \set{e}$}
    \EndIf
    \EndFor
    \Let{$i$}{$1$} \Comment{Size of candidates under consideration}
    \While{$C_{i} \neq \emptyset$}
    \Let{$C_{i+1}$}{$\emptyset$} \Comment{candidates of size $i+1$}
    \ForAll{$a, b \in C_{i}$ such that $\abs{a \cup b} = i+1$} \label{code:hbc:l:chooseab}
    \Let{$c$}{$a \cup b$} \label{code:hbc:l:unionab}
    \If{$c \setminus \set{e} \in C_{i}$ and $c \setminus \set{e}$ hits fewer sets than $c$ for all $e \in c$}
    \If{$c$ is a hitting set of $S$}
    \State \Output{$c$}
    \Else
    \Let{$C_{i+1}$}{$C_{i+1} \cup \set{c}$}
    \EndIf
    \EndIf
    \EndFor
    \Let{$i$}{$i+1$}
    \EndWhile
    \EndFunction
  \end{algorithmic}
\end{algorithm}

The authors claim a running time bound of $\compO{2^{x} \cdot y}$ where $x$ is the size of the largest hitting set and $y$ is the number of hitting sets of $S$.
However, Hagen shows in \cite{elbassioni2014} that this bound is incorrect.
He shows that \algname{MTMiner} is not output-polynomial and that its complexity is $n^{\compomega{\log \log n}}$, where $n = \abs{S} + y$.

It is possible to search only for MHSes of bounded cardinality with \algname{MTMiner} by discarding any candidate that is too large.
The second author and collaborators apply this approach as a \enquote{greedy algorithm} in \cite{veralicona2013} to study minimal interventions in a biochemical signalling network.
They take a different approach to element search than that in \cref{code:hbc:l:chooseab,code:hbc:l:unionab}; they instead loop over all candidate sets $a$ of a given size and consider $a \cup \set{e}$ for every element $e \notin a$ which does not form a singleton hitting set.
They also consider sets and elements in orders determined by a heuristic score called OCSANA to optimize the quality of approximate results in cases where complete enumeration is infeasible.

A C++ implementation of this algorithm by the authors of \cite{hebert2007} is available in the repository.

\subsubsection[MMCS and RS]{Murakami and Uno (2014)}
\label{s:alg-shd}
Murakami and Uno take a somewhat different approach in \cite{murakami2014} in two new algorithms.
We will refer to these algorithms as \algname{MMCS} and \algname{RS} (the names given to them by their authors).
Both rely on a crucial observation which makes possible efficient bottom-up searches for minimal hitting sets.

First, we require two definitions.
For a given family of sets $S$, a \emph{sub-MHS} is a set $M$ which is a subset of some MHS of $S$.
For a given set $E$ of elements of $S$, an element $e \in E$ is \emph{critical in $E$} if there is at least one set $s \in S$ which contains $e$ but no other elements of $E$.

Then we have the following proposition, appearing in various forms in cf.~\cite{hebert2007,murakami2014}:
\begin{proposition}
  A set $M$ of elements of a set family $S$ is a sub-MHS if and only if every $m \in M$ is critical in $M$.
  In this case, we say that $M$ satisfies the \emph{minimality condition}.
\end{proposition}

Thus, the MHSes of a set family are exactly the maximal element sets satisfying the minimality condition.
Both algorithms \algname{MMCS} and \algname{RS} proceed by building up sets that satisfy the minimality condition until they are hitting sets, making clever use of intermediate data structures to ensure that no redundant checks are performed.

Let $k = \norm{S}$ be the sum of the sizes of the sets in a set family $S$.
Then \algname{MMCS} runs in $\compO{k}$ time per iteration of its main loop, but the authors of \cite{murakami2014} do not give bounds for the number of iterations required.

For \algname{RS}, each iteration also takes $\compO{k}$ time, but the number of iterations can be bounded explicitly: it is $\compO{\sum y_{i}}$ for $y_{i}$ the number of MHSes of the subfamily $S_{\leq i} = \set{s_{1}, \dots, s_{i}}$.
Thus, the total running time is $\compO{k \cdot \sum y_{i}}$.

It is possible to search only for MHSes of bounded cardinality with \algname{MMCS} or \algname{RS} by simply discarding any candidate that is too large.
Furthermore, it is straightforward to parallelize the algorithm using the task model.
However, the \texttt{shd} program distributed by the authors of \cite{murakami2014} does not support either of these modes.

A C implementation of \algname{MMCS} and \algname{RS} by the authors of \cite{murakami2014} is available in the repository.
A C++ implementation of the parallel versions \algname{pMMCS} and \algname{pRS} which supports efficient enumeration of small MHSes is also included.

\subsection{Full cover approaches}
\label{s:alg-fullcover}
A fourth type of approach to computing the MHSes of a set family $S$ is to decompose the underlying elements into several subsets such that every set in $S$ lies entirely in one of them.
Formally, a \emph{full cover} of $S$ is another set family $C$ with the property that every $s \in S$ is a subset of some $c \in C$.

For any dual cover, we have the following decomposition result, given in cf.~\cite[Lemma 3]{boros2009}, which we express in the algebraic language of hypergraphs:

\begin{lemma}
  \label{lem:fullcover-decomp}
  Let $H$ be a simple hypergraph and let $C$ be a full cover of $H$.
  Then the transversal hypergraph $\Tr H$ of $H$ satisfies
  \begin{equation}
    \label{eq:fullcover-decomp}
    \Tr H = \bigwedge_{c \in C} \funceval{\Tr}{H_{c}}.
  \end{equation}
  where $\wedge$ is the wedge operation defined in \cref{def:hyp-join} and $H_{c}$ is the subhypergraph of $H$ containing only the edges that are subsets of $c$.
\end{lemma}

Given a set family $S$ and a full cover $C$ of $S$, we can use \cref{eq:fullcover-decomp} to break down the dualization computation into several independent computations which can be run in parallel, then merge the results using the hypergraph wedge operation.
(Each of these computations can in turn be decomposed recursively.)

There are two easy-to-find full covers of any set family $S$: the family $S$ itself and the singleton family $\set{\bigcup S}$.
The approaches given below use more refined full covers to ensure that the recursion of \cref{lem:fullcover-decomp} is efficient.

Pseudocode of this approach is given in \cref{code:fullcov} in the language of set families.

\begin{algorithm}
  \caption{Full cover algorithm\label{code:fullcov}}
  \begin{algorithmic}[1]
    \Require{A set family $S = \set{s_{1}, s_{2}, \dots, s_{n}}$ and a full cover $C$ of $S$}
    \Ensure{The set of MHSes of $S$}
    \Statex
    \Function{FullCoverDualize}{$S, C$}
    \ForAll{$c \in C$} \Comment{can be considered in parallel}
    \Let{$S_{c}$}{$\funceval{\min}{\setbuilder{s}{s \subseteq c}}$}
    \Let{$C_{c}$}{some full cover of $S_{c}$} \label{code:fullcov:l:chooseFC} \Comment{details vary by algorithm}
    \Let{$T_{c}$}{\Call{FullCoverDualize}{$S_{c}, C_{c}$}}
    \EndFor
    \Let{$T$}{$\funceval{\min}{\setbuilder{\bigcup_{c \in C} t_{c}}{t_{c} \in T_{c}}}$} \label{code:fullcov:l:wedge} \Comment{hypergraph wedge}
    \State \Return{$T$}
    \EndFunction
  \end{algorithmic}
\end{algorithm}

Of course, the efficiency of this algorithm depends on the choice of full cover in \cref{code:fullcov:l:chooseFC}.
In particular, the procedure to choose this full cover $C$ should have three properties:
\begin{enumerate}
\item
  $C$ should have many components, to spread the load over many processors;

\item
  the individual computations $\Call{FullCoverDualize}{S_{c}, C_{c}}$ should be substantially smaller in scale than the full computation, so no one processor has too much work to do; and

\item
  the merge operation in \cref{code:fullcov:l:wedge} (and in particular the minimization step) should not be too complex, so the sequential part of the algorithm does not dominate the running time.
\end{enumerate}

Several published algorithms fit into this scheme; they differ primarily in how they approach the construction of $C$.

\subsubsection[KBEG]{Khachiyan \textit{et al.} (2007)}
\label{s:alg-kbeg}
Khachiyan et al.\ introduced the full cover decomposition approach in \cite{khachiyan2007global,khachiyan2007computing}.
They focus on hypergraphs with a curious property: the restriction of the hypergraph to any vertex subset $V'$ admits a full cover with the property that each covering edge has size less than $\pbrac*{1 - \epsilon} \abs{V'}$ for a fixed threshold parameter $0 < \epsilon < 1$.
They show that using such a collection of full covers in \cref{lem:fullcover-decomp} yields an efficient recursive algorithm.
They then are able to show that this procedure runs in polylogarithmic time on polynomially many processors, with coefficients determined by the value of $\epsilon$.
We will refer to this algorithm as \algname{pKBEG} (for the Parallel algorithm of Khachiyan, Boros, Elbassioni, and Gurvich).

Of course, this analysis only applies if such a family of full covers can be found.
They demonstrate constructions (and give explicit values of $\epsilon$) for several important families: hypergraphs with bounded edge size (\enquote{dimension}), bounded \enquote{dual-conformality} (a condition related to intersections in the transversal), or bounded edge-transversal intersection size.

The authors are not aware of a public implementation of this algorithm.
Since it does not apply in generality, we will not study it in \cref{s:experiment}.

\subsubsection[pELB]{Elbassioni (2008)}
\label{s:alg-pelb}
Following up on \cite{khachiyan2007global,khachiyan2007computing}, Elbassioni presents in \cite{elbassioni2008complexity} two parallel decomposition approaches for the transversal recognition problem.
The first is essentially a rearrangement of \algname{FK-B} to make the search tree broader and shallower so parallel computation is efficient.
The second is a variant of a full-cover decomposition algorithm; given a transversal $T$ of a hypergraph $H$ with vertex set $V$, it uses
\begin{equation}
  \label{eq:fullcover-hs}
  C(T) = \setbuilder{V \setminus \cbrac{i}}{i \in T} \cup \cbrac{T}
\end{equation}
as a full cover of $\Tr H$ to decompose the problem.
(He also incorporates a special divide-and-conquer case similar to the \algname{FK} algorithms under certain circumstances.)
We will refer to this algorithm as \algname{pELB} (for the Parallel algorithm of Elbassioni).

Elbassioni shows that this algorithm runs in polylogarithmic time on quasipolynomially many processors in polynomial space for any hypergraph; in particular, letting $n$ be the number of vertices, $x$ be the number of edges of $H$, and $y$ be the size of $\Tr H$, the running time is bounded by both $n^{2} x^{\compo{\log y}}$ and $n^{2} y^{\compo{\log x}}$, so any asymmetry in the sizes of $H$ and $\Tr H$ reduces the runtime.
(The exact bounds are cumbersome to state but may be found in \cite{elbassioni2008complexity}.)

The authors are not aware of any public implementation of this algorithm.

\subsubsection[pBM]{Boros and Makno (2009) }
\label{s:alg-pbm}
Boros and Makino present in \cite{boros2009} a full cover algorithm which improves on the asymptotic complexity bounds of \cite{elbassioni2008complexity} for transversal recognition.
To do this, they introduce another full cover in addition to that in \cref{eq:fullcover-hs}, which they incorporate into an \algname{FK}-like recursive duality-testing framework.
Fix a hypergraph $H$ and an edge $e \in H$; then
\begin{equation}
  \label{eq:fullcover-edge}
  C(e) = \setbuilder{\pbrac*{V \setminus f} \cup \cbrac{i}}{f \in H, i \in f \cap e}
\end{equation}
is a full cover of $\Tr A$.
By carefully choosing when to use a full cover from \cref{eq:fullcover-hs} or \cref{eq:fullcover-edge}, Boros and Makino are able to obtain very strong bounds on parallel runtime.
We will refer to this algorithm as \algname{pBM} (for the Parallel algorithm of Boros and Makino).

Fix a hypergraph $H$ with $n$ vertices and $x$ edges for which $\Tr H$ has $y$ edges.
Then \algname{pBM} runs in $\compO{\log n + \log x \log y}$ time using $\compO{n x y^{1 + \log x}}$ processors.

A C++ implementation of this algorithm for MHS generation, written by the first author, is available in the repository.

\subsection{Other}
\label{s:alg-other}
Some authors have used approaches that translate the MHS generation problem into other domains for which specialized algorithms already exist.
We outline these below.

\subsubsection{Primary decomposition of squarefree monomial ideals}
\label{s:alg-primdecomp}
The MHS generation problem can be translated into a problem in computational algebra.
Fix a set family $S = \set{s_{1}, s_{2}, \dots, s_{n}}$ with underlying element set $E = \bigcup_{i} s_{i} = \set{e_{1}, \dots, e_{m}}$
To each element $e_{i}$ associate a variable $x_{i}$ in a polynomial ring over $\mathbf{Q}$.
To each set $s_{i}$, associate a monomial $m_{i} = \prod_{e_{j} \in s_{i}} x_{j}$.
(For example, the set $\set{1, 2, 5}$ becomes the monomial $x_{1} x_{2} x_{5}$).
We can then construct a monomial ideal $I_{S}$ generated by the monimials $m_{s}$, which encodes the set family algebraically.
By construction, $I_{S}$ is squarefree.
It then turns out that the generators of the associated primes of $I_{S}$ correspond exactly to the minimal hitting sets of $H$.
We will refer to this approach as \algname{PrimDecomp}.

This approach was used by Jarrah \textit{et al.} in \cite{jarrah2007} for an application in computational biology.
They calculate the associated primes of $I_{S}$ using Alexander duality \cite{miller1998} as provided in Macaulay2 \cite{macaulay2}.

A container which uses Macaulay2 to perform this calculation is provided in the repository.

\subsubsection{Integer programming}
\label{s:alg-intprog}
The MHS generation problem can be interpreted as an integer programming problem.
Fix a set family $S = \set{s_{1}, s_{2}, \dots, s_{n}}$ with underlying element set $E = \bigcup_{i} s_{i} = \set{e_{1}, \dots, e_{m}}$
We declare $n$ variables $x_{i}$, each of which may take values from $\set{0, 1}$.
A subset $T$ of the vertices then corresponds to an assignment $\mathbf{x}$ of the $x$-variables.
For each set $s_{i}$, we impose a constraint $\sum_{e_{j} \in s} x_{j} \geq 1$; an assignment $\mathbf{x}$ corresponds to a hitting set if it satisfies all these constraints.
Enumeration of inclusion-minimal assignments that satisfy the constraints is then exactly the MHS generation problem.
(Indeed, it was shown by Boros \textit{et al.} in \cite{boros2002dual} that MHS generation is equivalent to the general problem of enumerating minimal solutions to the linear system $Ax = b$ for $0 \leq x \leq c$ where $A$ is a binary matrix, $x$ is a binary vector, and $b$ and $c$ are all-ones vectors.)
We will refer to this approach as \algname{IntProg}.

Because linear programming solvers are so diverse and many widely-used ones are proprietary, we do not provide an implementation of this approach.

\subsection{Feature comparison}
\label{s:alg-feature}
We summarize in \cref{tab:alg-feature} the salient features of the algorithms introduced in \cref{s:algorithms}.

\ctable[
caption = Feature comparison of MHS generation algorithms,
label = tab:alg-feature,
pos = htbp,
]{*{3}{l} *{4}{c}}{
\tnote[a]{
  \checkmark indicates that an algorithm is evaluated in \cref{s:experiment}.
}
\tnote[b]{Indicates that an algorithm can generate only small MHSs to save time}
\tnote[c]{Separate parallel implementation by the first author}
}{
    \toprule
    & Algorithm & Domain & Published & Eval.\tmark[a] & Parallel & Cutoff\tmark[b] \\
    \midrule
    \rowlabel{Set iteration}
    & \algname{Berge} (\S \ref{s:alg-berge}) & Hypergraphs & 1984 & \checkmark & & \checkmark \\
    & \algname{HS-DAG} (\S \ref{s:alg-hsdag}) & Fault diagnosis & 1989 &\checkmark & & \checkmark \\
    & \algname{HST} (\S \ref{s:alg-hst}) & Fault diagnosis & 2001 & \checkmark & & \checkmark \\
    & \algname{BMR} (\S \ref{s:alg-bmr}) & Data mining & 2003 & \checkmark & & \\
    & \algname{DL} (\S \ref{s:alg-dl}) & Data mining & 2005 & \checkmark & & \\
    & \algname{KS} (\S \ref{s:alg-ks}) & Hypergraphs & 2005 & \checkmark & & \\

    \\\rowlabel{Divide and conquer}
    & \algname{FK-A} (\S \ref{s:alg-fk}) & Boolean algebra & 1996 & \checkmark & & \\
    & \algname{FK-B} (\S \ref{s:alg-fk}) & Boolean algebra & 1996 & & & \\
    & \algname{BOOL} (\S \ref{s:alg-bool}) & Fault diagnosis & 2003 & \checkmark & & \checkmark \\
    & \algname{STACCATO} (\S \ref{s:alg-staccato}) & Fault diagnosis & 2009 & \checkmark & & \checkmark \\
    & \algname{ParTran} (\S \ref{s:alg-partran}) & Poset theory & 2010 & \checkmark & \checkmark & \\
    & \algname{Knuth} (\S \ref{s:alg-knuth}) & Boolean algebra & 2011 & \checkmark & & \\
    & \algname{HTC-BDD} (\S \ref{s:alg-htcbdd}) & Boolean algebra & 2013 & \checkmark & & \\
    & \algname{\MHStwo} (\S \ref{s:alg-mhs2}) & Fault diagnosis & 2014 & \checkmark & \checkmark & \checkmark \\

    \\\rowlabel{MHS buildup}
    & \algname{MTMiner} (\S \ref{s:alg-hbc}) & Data mining & 2007 & \checkmark & & \checkmark \\
    & \algname{MMCS} (\S \ref{s:alg-shd}) & Hypergraphs & 2014 & \checkmark & \checkmark\tmark[c] & \checkmark \\
    & \algname{RS} (\S \ref{s:alg-shd}) & Hypergraphs & 2014 & \checkmark & \checkmark\tmark[c] & \checkmark \\

    \\\rowlabel{Full covers}
    & \algname{pKBEG} (\S \ref{s:alg-kbeg}) & Boolean algebra & 2007 & & \checkmark & \\
    & \algname{pELB} (\S \ref{s:alg-pelb}) & Boolean algebra & 2008 & & \checkmark & \\
    & \algname{pBM} (\S \ref{s:alg-pbm}) & Boolean algebra & 2009 & \checkmark & \checkmark & \\

    \\\rowlabel{Other}
    & \algname{PrimDecomp} (\S \ref{s:alg-primdecomp}) & Comp.~algebra & 2007 & \checkmark & & \\
    & \algname{IntProg} (\S \ref{s:alg-intprog}) & Optimization & -- & & & \\
    \bottomrule
}

\subsection{Algorithm miscellany}
\label{s:alg-misc}

A genetic algorithm for finding many (but not necessarily all) small (but not necessarily minimal) hitting sets is studied by Li and Yunfei in \cite{li2002}.
Vinterbo and \O{}hrn study in \cite{vinterbo2000} the more refined problem of finding weighted $r$-approximate hitting sets, which are sets which hit some fraction $0 \leq r \leq 1$ of the target sets according to assigned weights; they also apply a genetic algorithm with promising results.

Jelassi \textit{et al.} consider the efficacy of pre-processing methods in \cite{jelassi2015}.
They find that, for many common classes of set families, it is worthwhile to compute from the family $S$ a new family $S'$ which combines elements which occur only in the same sets into so-called \emph{generalized nodes}.
(This optimization was also used by Kavvadias and Stavropoulos in \cite{kavvadias2005} for their algorithm \algname{KS}.)
Their algorithm \algname{Irred-Engine} performs this preprocessing, applies a known MHS algorithm (in their case, \algname{MMCS} from \cite{murakami2014}) to the resulting family $S'$, and then expands the results into MHSes for the original $S$.
We will not study this approach separately here, but it may be of interest for applications where many vertices may be redundant.

\section{Time-performance comparison of the algorithms}
\label{s:experiment}
Numerous previous papers have included experimental comparisons of some algorithms, including \cite{murakami2014,leiserson2010,kavvadias2005,hebert2007,hagen2009fk,dong2005,cardoso2014,bailey2003}.
However, we find that there is need for a new, comprehensive survey for several reasons:
\begin{enumerate}
\item
  Each published comparison involves only a few algorithms, and differences in data sets and environment make the results incompatible.
  Thus, it is not possible to assemble a systematic overview of the relative performance of these algorithms.

\item
  Many existing comparisons overlook published algorithms in domains far from the authors' experience.
  An algorithm designed for monotone dualization may prove to be useful for data mining, for example, but authors in that field may be unaware of it due to translational issues in the literature.

\item
  Most existing comparisons are not published alongside working code and do not provide methodological details so that the results can be reproduced or extended.
  (Murakami and Uno's work in \cite{murakami2014} is a notable exception, and indeed their publicly-available implementations are used for several algorithms here.)
\end{enumerate}

\subsection{Methodology}
\label{s:methodology}
We have assembled a repository of software implementations of existing algorithms.
Each is wrapped in a Docker container using the Algorun framework and using standardized JSON formats for input and output.
Details, code, and containers are available from \url{https://github.com/VeraLiconaResearchGroup/MHSGenerationAlgorithms}, including complete instructions for reproducing the experimental environment and running new experiments.
These containers are easy to deploy on any computer supporting the Docker container environment; they do not require compiling any code or downloading libraries.
Interested readers are encouraged to run similar experiments on their own data sets.

We have run each implemented algorithm on a variety of input set families (discussed in detail in \cref{s:datasets}).
Each was allowed to run for up to one hour (\SI{3600}{\second}) before termination; at least one algorithm ran to termination on every data set with this timeout.
Algorithms which did not time out were run a total of three times and the median runtime used for analysis, presented in \cref{s:results}.
Algorithms which support cutoff enumeration (that is, finding only hitting sets of size up to some fixed $c$) were run with $c = 5$, $7$, and $10$ as well as full enumeration.
Algorithms which support multiple threads were run with $t = 1$, $2$, $4$, $6$, $8$, $12$, and $16$ threads.
All experiments were performed on a workstation with an Intel Xeon E5-2630v3 processor with eight cores at \SI{2.4}{\giga\hertz} (with Hyperthreading enabled, allowing 16 concurrent threads) and \SI{32}{\giga\byte} of ECC DDR4 RAM.

In all cases, the generated hitting sets were compared to ensure that the algorithms were running correctly.
This revealed errors in several published implementations, which are discussed in \cref{s:algorithms,tab:alg-feature}.
In cases where an algorithm's results are only slightly incorrect, we have included its benchmark timing in the results below, since we believe these still give a useful impression of the relative performances of these algorithms.

\subsection{Data sets used for time-performance comparison}
\label{s:datasets}
We apply each algorithm to a variety of set families derived from real-world data.
We briefly discuss each data set here.
We have focused on data sets that provide large, heterogenous set families, since these cases highlight the performance differences among algorithms; for smaller families, the differences may be negligible in practice.

\begin{description}[font=\datasetname,style=nextline]
\item[accident]
  Anonymized information about several hundred thousand accidents in Flanders during the period 1991--2000.
  Originally published in \cite{geurts2003}.
  Converted by the authors of \cite{murakami2014} into a set family whose sets are the complements of maximal frequent itemsets with specified threshold $1000 \theta$ for $\theta \in \set{70, 90, 110, 130, 150, 200}$; MHSes of this set family then correspond to minimal infrequent itemsets.
  All of the set families have 441 underlying elements; numbers of sets range from 81 (for $\theta = 200$) to 10968 (for $t = 70$).
  This formulation was downloaded from \cite{dual-rep}.

\item[ecoli]
  Metabolic reaction networks from \textit{E.~coli}.
  Reaction networks for producing acetate, glucose, glycerol, and succinate, along with the combined network, were analyzed to find their \enquote{elementary modes} using \algname{Metatool} \cite{vonkamp2006}, which are given as set families.
  MHSes of these set families correspond to \enquote{minimal cut sets} of the original networks, which are of interest in studying and controlling these networks.
  Statistics for these set families are given in \cref{tab:ecoli}.

  \ctable[
  caption = Statistics for \textit{E.~coli} network data sets,
  label = tab:ecoli,
  pos = htbp,
  ]{c *{3}{r}}{}{
    \toprule
    Network & elements & sets & avg.~set size \\
    \midrule
    Acetate & 103 & 266 & 23.7 \\
    Glucose & 104 & 6387 & 30.4 \\
    Glycerol & 105 & 2128 & 27.2 \\
    Succinate & 103 & 932 & 22.3 \\
    Combined & 109 & 27503 & 30.6 \\
    \bottomrule
  }

\item[ocsana]
  Interventions in cell signalling networks.
  Two cell signalling networks (EGFR from \cite{samaga2009} and HER2+ from \cite{veralicona2012}) were analyzed to find their \enquote{elementary pathways} using \algname{OCSANA}, which are given as set families.
  MHSes of these set families correspond to \enquote{optimal combinations of interventions} in the original networks, which are of interest in studying and controlling these networks.
  Each network  has been preprocessed to find these elementary pathways using three different algorithms of increasing resolution: shortest paths only (SHORT), including \enquote{suboptimal} paths (SUB), and including all paths up to length 20 (ALL).
  This results in six set families.
  Statistics for these families are given in \cref{tab:ocsana}.

  \ctable[
  caption = Statistics for OCSANA network data sets,
  label = tab:ocsana,
  pos = htbp,
  ]{*{2}{c} *{3}{r}}{}{
    \toprule
    Network & Method & elements & sets & avg.~set size \\
    \midrule
    EGFR & SHORT & 49 & 125 & 8.9 \\
    & SUB &  55 & 234 & 9.9 \\
    & ALL & 63 & 11050 & 16.5 \\
    HER2+ & SHORT & 122 & 534 & 15.2 \\
    & SUB & 171 & 2538 & 20.3 \\
    & ALL & 318 & 69805 & 19.1 \\
    \bottomrule
  }

  These set families demonstrate particularly effectively the problem of combinatorial explosion in MHS generation.
  For example, the \datasetname{HER2+.SHORT} set family has just 122 underlying elements and 534 sets, but we have computed that it has 128833310 MHSes.
  Even storing the collection of MHSes in memory is difficult because of its size.
  As a result, none of the algorithms tested were able to complete the full enumeration of MHSes for any of the \datasetname{HER2+} data sets.
  However, the cutoff enumeration is much more manageable; for example, \datasetname{HER2+.SHORT} has just 26436 MHSes of size $c \leq 7$, which are found easily by several of the algorithms under study.
\end{description}

\subsection{Results}
\label{s:results}

We present below the results of the benchmarking experiments on the data sets described in \cref{s:datasets}.
All experiments were performed on a workstation with an Intel Xeon E5-2630v3 processor with eight cores at \SI{2.4}{\giga\hertz} (with Hyperthreading enabled, allowing 16 concurrent threads) and \SI{32}{\giga\byte} of ECC DDR4 RAM.
Each algorithm was allowed to run for up to \SI{3600}{\second}; algorithms that did not complete in this time are marked with \algtimeout, while algorithms that crashed due to memory exhaustion are marked with \algmemexhaust.

\subsubsection{Full enumeration}
We first consider the general problem of enumerating all MHSes of a given set family.
Timing results for the full enumeration cases are given in \cref{tab:results-full.accident,tab:results-full.ocsana.egfr,tab:results-full.ecoli}, with algorithms sorted in approximately increasing order of speed.

\begin{sidewaystable}
  \centering
  \begin{filecontents*}{data/full.accident.csv}
Input,ac-200k,ac-150k,ac-130k,ac-110k,ac-090k,ac-070k,ac-050k,ac-030k
mmcs,0.0032000542,0.0104990005,0.0210459232,0.0588648319,0.2309238911,0.664634943,2.2774951458,20.2630419731
rs,0.0025610924,0.0102851391,0.0285031796,0.0599548817,0.2299098969,0.6477060318,2.2688901424,19.8359658718
pMMCS,0.0066819191,0.0167818069,0.0332651138,0.0582370758,0.2237110138,0.5588040352,1.8187801838,17.3169109821
pRS,0.0065360069,0.0193319321,0.0347390175,0.0740330219,0.2640390396,0.7515201569,3.0539751053,46.8105640411
mtminer,0.0050799847,0.0174760818,0.044178009,0.0847370625,0.32030797,0.9351730347,3.4543049336,28.5911989212
bmr,0.0074079037,0.0500731468,0.0973360538,0.1773929596,1.3997290134,3.4360001087,13.6017508507,108.559787989
htcbdd,0.3805930614,0.3938698769,0.4582941532,0.4429850578,0.5921599865,0.8274431229,{\algmemexhaust},{\algmemexhaust}
knuth,0.3212840557,0.3479261398,0.3790578842,0.4067909718,0.7314519882,1.3443789482,{\algmemexhaust},{\algmemexhaust}
mhs2,0.0069169998,0.0428099632,0.1204550266,0.2850379944,2.6742460728,14.7229127884,125.856606007,{\algtimeout}
dl,0.0083041191,0.0727829933,0.227134943,0.6979169846,3.0621240139,17.3767080307,146.441184044,2763.5942688
fka-begk,0.2642128468,1.1183569431,2.8802320957,7.0770480633,64.2280228138,321.597411871,2271.21517396,{\algmemexhaust}
bool,0.0366580486,0.2387390137,1.2591269016,3.0752449036,111.487586975,493.715492964,{\algtimeout},{\algtimeout}
hst,0.1154141426,2.7341821194,12.2646808624,54.7993819714,283.953449011,1855.83450913,{\algtimeout},{\algtimeout}
primdecomp,0.533921957,1.1885259152,3.6551799774,7.3798480034,272.63094902,888.387526989,{\algtimeout},{\algmemexhaust}
hsdag,0.2617220879,2.1069099903,11.9114890099,41.3045539856,2344.98416996,{\algtimeout},{\algtimeout},{\algtimeout}
berge,0.3846600056,6.7094581127,52.7100739479,289.998375893,{\algtimeout},{\algtimeout},{\algtimeout},{\algtimeout}
partran,0.8417270184,16.4567639828,167.949976921,727.055366039,{\algtimeout},{\algtimeout},{\algtimeout},{\algtimeout}
pbm,3.2015240192,1758.70500898,{\algtimeout},{\algtimeout},{\algtimeout},{\algtimeout},{\algtimeout},{\algtimeout}
staccato,45.5897278786,{\algtimeout},{\algtimeout},{\algtimeout},{\algtimeout},{\algtimeout},{\algtimeout},{\algtimeout}
verts,{\algcount{64}},{\algcount{64}},{\algcount{81}},{\algcount{81}},{\algcount{335}},{\algcount{336}},{\algcount{336}},{\algcount{442}}
edges,{\algcount{81}},{\algcount{447}},{\algcount{990}},{\algcount{2000}},{\algcount{4322}},{\algcount{10968}},{\algcount{32207}},{\algcount{135439}}
MHSes,{\algcount{253}},{\algcount{1039}},{\algcount{1916}},{\algcount{3547}},{\algcount{7617}},{\algcount{17486}},{\algcount{47137}},{\algcount{185218}}
\end{filecontents*}
\pgfplotstabletypeset[benchmarktable =
  {
    Algorithm & \multicolumn{8}{c}{\datasetname{accident} (threshold $\theta$ in thousands of incidents, smaller $\theta$ gives larger data set)} \\
    \cmidrule(lr){2-9}
    & {$\theta = 200$} & {$\theta = 150$} & {$\theta = 130$} & {$\theta = 110$} & {$\theta = 90$} & {$\theta = 70$} & {$\theta = 50$} & {$\theta = 30$} \\
  }
  ]{data/full.accident.csv}
  \algcaption{accident}{with various cutoff values $\theta$}
  \label{tab:results-full.accident}
\end{sidewaystable}

\begin{table}
  \centering
  \begin{filecontents*}{data/full.ocsana.egfr.csv}
Input,EGFR.short,EGFR.sub,EGFR.all
mmcs,0.0075259209,0.0484509468,1.3401658535
rs,0.0078449249,0.0469520092,1.3330929279
pMMCS,0.0829911232,0.4611470699,3.2097182274
pRS,0.0447130203,0.3681540489,4.6736309528
mtminer,{\algmemexhaust},{\algmemexhaust},{\algmemexhaust}
bmr,0.2199039459,2.9872760773,22.2575871944
htcbdd,0.4024238586,0.5038719177,0.5073068142
knuth,0.3672778606,{\algmemexhaust},{\algmemexhaust}
mhs2,{\algtimeout},{\algtimeout},{\algmemexhaust}
dl,0.1448779106,4.2471909523,39.660533905
fka-begk,10.3938820362,152.44046402,705.065476894
bool,12.1779139042,3115.05892396,2865.42398691
hst,{\algtimeout},{\algtimeout},{\algmemexhaust}
primdecomp,1.0125341415,10.0030720234,13.9054100513
hsdag,{\algtimeout},{\algtimeout},{\algmemexhaust}
berge,0.3051240444,9.8914821148,2065.66590714
partran,289.409902811,{\algtimeout},{\algtimeout}
pbm,1544.0026021,{\algtimeout},{\algtimeout}
staccato,{\algtimeout},{\algtimeout},{\algtimeout}
verts,{\algcount{51}},{\algcount{57}},{\algcount{65}}
edges,{\algcount{125}},{\algcount{234}},{\algcount{11050}}
MHSes,{\algcount{1340}},{\algcount{11765}},{\algcount{13116}}
\end{filecontents*}
\pgfplotstabletypeset[benchmarktable =
  {
    Algorithm & \multicolumn{3}{c}{\datasetname{EGFR}} \\
    \cmidrule(lr){2-4}
    & {\datasetname{short}} & {\datasetname{sub}} & {\datasetname{all}} \\
  }
  ]{data/full.ocsana.egfr.csv}
  \algcaption{ocsana-egfr}{with the path-finding strategies \datasetname{short}, \datasetname{sub}, and \datasetname{all}}
  \label{tab:results-full.ocsana.egfr}
\end{table}

\begin{table}
  \centering
  \begin{filecontents*}{data/full.ecoli.csv}
Input,ecoli-acetate,ecoli-succinate,ecoli-glycerol,ecoli-glucose,ecoli-combined
mmcs,0.0520939827,0.8557710648,2.5119080544,4.7186620235,157.081720114
rs,0.0557098389,0.7507228851,2.4607579708,4.5063240528,155.629385948
pMMCS,0.0677211285,0.4626791477,1.1346538067,4.4252479076,202.647791147
pRS,0.3285188675,3.9492621422,15.4907209873,26.3078010082,1989.8463459
mtminer,149.855538845,682.336236954,{\algmemexhaust},{\algmemexhaust},{\algmemexhaust}
bmr,0.5624530315,12.1329979897,33.1988937855,26.7359220982,2209.63064408
htcbdd,0.4686000347,0.553278923,0.6877930164,0.8473651409,14.1573348045
knuth,{\algmemexhaust},{\algmemexhaust},{\algmemexhaust},{\algmemexhaust},{\algmemexhaust}
mhs2,140.503093004,737.680742025,2447.52871799,{\algtimeout},{\algmemexhaust}
dl,3.1477489471,225.505749941,1771.77100301,1749.87818503,{\algmemexhaust}
fka-begk,25.8001179695,514.179192066,1421.67719102,{\algmemexhaust},{\algmemexhaust}
bool.iterative,5.1064457893,438.794944048,{\algtimeout},{\algtimeout},{\algmemexhaust}
hst,{\algtimeout},{\algtimeout},{\algtimeout},{\algtimeout},{\algmemexhaust}
primdecomp,3.4635369778,22.4183170795,53.2389957905,62.5651910305,{\algmemexhaust}
hsdag,2945.18629408,{\algtimeout},{\algtimeout},{\algtimeout},{\algmemexhaust}
berge,52.8984558582,{\algtimeout},{\algtimeout},{\algtimeout},{\algtimeout}
partran,{\algtimeout},{\algtimeout},{\algtimeout},{\algtimeout},{\algmemexhaust}
pbm,{\algtimeout},{\algtimeout},{\algtimeout},{\algtimeout},{\algtimeout}
staccato,{\algtimeout},{\algtimeout},{\algtimeout},{\algtimeout},{\algmemexhaust}
verts,{\algcount{103}},{\algcount{103}},{\algcount{105}},{\algcount{104}},{\algcount{109}}
edges,{\algcount{266}},{\algcount{932}},{\algcount{2128}},{\algcount{6387}},{\algcount{27503}}
MHSes,{\algcount{3363}},{\algcount{14136}},{\algcount{25619}},{\algcount{21001}},{\algcount{275914}}
\end{filecontents*}
\pgfplotstabletypeset[benchmarktable =
  {
    Algorithm & \multicolumn{5}{c}{\datasetname{ecoli}} \\
    \cmidrule(lr){2-6}
    & {\datasetname{acetate}} & {\datasetname{succinate}} & {\datasetname{glycerol}} & {\datasetname{glucose}} & {\datasetname{combined}} \\
  }
  ]{data/full.ecoli.csv}
  \algcaption{ecoli}{networks}
  \label{tab:results-full.ecoli}
\end{table}

\subsubsection{Multithreaded full enumeration}
Although many of the published algorithms are serial, a few can be parallelized.
For the algorithms for which multithreaded implementations were available, we have run tests with $t \in \set{1, 2, 4, 6, 8, 12, 16}$ threads on our workstation with eight true cores and Hyperthreading support.
Timing results for selected full enumeration cases with various numbers of threads are shown in \cref{tab:results-thread-ec-acetate,tab:results-thread-ec-combined}.

\begin{table}
  \centering
  \begin{filecontents*}{data/threaded.ecoli.acetate.csv}
Input,ecoli.acetate-t1,ecoli.acetate-t2,ecoli.acetate-t4,ecoli.acetate-t6,ecoli.acetate-t8,ecoli.acetate-t12,ecoli.acetate-t16
pMMCS,0.0677211285,0.0608029366,0.0405209064,0.0430030823,0.0431718826,0.0566949844,0.0677089691
pRS,0.3285188675,0.2342939377,0.1969840527,0.1427381039,0.1257557869,0.1429469585,0.1249918938
mhs2,140.503093004,90.9883048534,62.5167350769,61.3615159988,60.5991849899,60.8542349339,58.9037649632
pbm,{\algtimeout},{\algtimeout},{\algtimeout},{\algtimeout},{\algtimeout},{\algtimeout},{\algtimeout}
\end{filecontents*}
\pgfplotstabletypeset[benchmarktable =
  {
    Algorithm & \multicolumn{7}{c}{\datasetname{ecoli-acetate}} \\
    \cmidrule(lr){2-8}
    & {$t = 1$} & {$t = 2$} & {$t = 4$} & {$t = 6$} & {$t = 8$} & {$t = 12$} & {$t = 16$} \\
  }]{data/threaded.ecoli.acetate.csv}
  \algcaptionthread{ecoli-acetate}{}
  \label{tab:results-thread-ec-acetate}
\end{table}

\begin{table}
  \centering
  \begin{filecontents*}{data/threaded.ecoli.combined.csv}
Input,ecoli.combined-t1,ecoli.combined-t2,ecoli.combined-t4,ecoli.combined-t6,ecoli.combined-t8,ecoli.combined-t12,ecoli.combined-t16
pMMCS,202.647791147,118.702360153,72.0508308411,60.0077619553,50.7631340027,49.160820961,51.5810508728
pRS,1989.8463459,1072.64463019,591.160926104,431.817724943,364.220141888,348.7759161,334.610668182
mhs2,{\algmemexhaust},{\algmemexhaust},{\algmemexhaust},{\algmemexhaust},{\algmemexhaust},{\algmemexhaust},{\algmemexhaust}
pbm,{\algtimeout},{\algtimeout},{\algtimeout},{\algtimeout},{\algtimeout},{\algtimeout},{\algtimeout}
\end{filecontents*}
\pgfplotstabletypeset[benchmarktable =
  {
    Algorithm & \multicolumn{7}{c}{\datasetname{ecoli-combined}} \\
    \cmidrule(lr){2-8}
    & {$t = 1$} & {$t = 2$} & {$t = 4$} & {$t = 6$} & {$t = 8$} & {$t = 12$} & {$t = 16$} \\
  }]{data/threaded.ecoli.combined.csv}
  \algcaptionthread{ecoli-combined}{}
  \label{tab:results-thread-ec-combined}
\end{table}

\subsubsection{Cutoff enumeration}
In many applications, only small MHSes are relevant.
We consider here the enumeration of MHSes of size no greater than some \enquote{cutoff} $c$; we have run benchmarks for $c \in \set{5, 7, 10}$ using the algorithms which support cutoff mode.
Timing results for selected cutoff enumeration cases are given in \cref{tab:results-cutoff.ocsana,tab:results-cutoff.ecoli}.

\begin{sidewaystable}
  \centering
  \begin{filecontents*}{data/cutoff.ocsana.csv}
Input,HER2.short-c5,HER2.short-c7,HER2.short-c10,HER2.all-c5,HER2.all-c7,HER2.all-c10
pMMCS,0.0958721638,1.0769441128,{\algmemexhaust},5.7063500881,64.8871798515,{\algmemexhaust}
pRS,0.083261013,2.2265539169,{\algmemexhaust},1.8801798821,94.6899380684,{\algmemexhaust}
mhs2,649.805147886,{\algtimeout},{\algtimeout},{\algtimeout},{\algtimeout},{\algtimeout}
bool,1.2323849201,42.1837439537,{\algmemexhaust},201.705332041,{\algmemexhaust},{\algmemexhaust}
hst,26.9258289337,{\algtimeout},{\algtimeout},10.0001890659,{\algtimeout},{\algtimeout}
hsdag,0.8417439461,1218.22850919,{\algtimeout},2.205862999,666.131658792,{\algtimeout}
berge,5.1779839993,{\algtimeout},{\algtimeout},42.0164470673,{\algtimeout},{\algtimeout}
staccato,{\algtimeout},{\algtimeout},{\algtimeout},{\algtimeout},{\algtimeout},{\algtimeout}
verts,{\algcount{124}},{\algcount{124}},{\algcount{124}},{\algcount{320}},{\algcount{320}},{\algcount{320}}
edges,{\algcount{534}},{\algcount{534}},{\algcount{534}},{\algcount{69805}},{\algcount{69805}},{\algcount{69805}}
MHSes,{\algcount{88}},{\algcount{26436}},{\algcount{8744333}},{\algcount{40}},{\algcount{1892}},{\algcount{2853026}}
\end{filecontents*}
\pgfplotstabletypeset[benchmarktable =
  {
    Algorithm & \multicolumn{6}{c}{\datasetname{HER2}} \\
    & \multicolumn{3}{c}{\datasetname{short}} & \multicolumn{3}{c}{\datasetname{all}} \\
    \cmidrule(lr){2-4} \cmidrule{5-7}
    & {$c = 5$} & {$c = 7$} & {$c = 10$} & {$c = 5$} & {$c = 7$} & {$c = 10$} \\
  }]{data/cutoff.ocsana.csv}
  \algcaptioncutoff{ocsana-HER2+}{with path-finding strategies \datasetname{short} and \datasetname{all}}
  \label{tab:results-cutoff.ocsana}
\end{sidewaystable}

\begin{sidewaystable}
  \centering
  \begin{filecontents*}{data/cutoff.ecoli.csv}
Input,ecoli.acetate-c5,ecoli.acetate-c7,ecoli.acetate-c10,ecoli.combined-c5,ecoli.combined-c7,ecoli.combined-c10
pMMCS,0.0093581676,0.0123698711,0.0314021111,0.4017221928,1.8952140808,17.935033083
pRS,0.0092208385,0.0279791355,0.0997531414,0.2711970806,4.1242699623,118.489578962
mhs2,1.3181960583,12.1392550468,78.9267179966,870.358914852,{\algmemexhaust},{\algmemexhaust}
bool,0.0228819847,0.0867910385,0.7313859463,21.4547131062,198.682071924,{\algmemexhaust}
hst,0.016215086,8.2261669636,{\algtimeout},8.9210209847,{\algmemexhaust},{\algmemexhaust}
hsdag,0.0232460499,0.5467560291,45.9378409386,6.9582281113,546.502822876,{\algmemexhaust}
berge,0.2022259235,0.9275500774,3.799957037,280.453153849,{\algtimeout},{\algtimeout}
staccato,{\algtimeout},{\algtimeout},{\algtimeout},{\algmemexhaust},{\algmemexhaust},{\algmemexhaust}
verts,{\algcount{103}},{\algcount{103}},{\algcount{103}},{\algcount{110}},{\algcount{110}},{\algcount{110}}
edges,{\algcount{266}},{\algcount{266}},{\algcount{266}},{\algcount{27503}},{\algcount{27503}},{\algcount{27503}}
MHSes,{\algcount{39}},{\algcount{195}},{\algcount{735}},{\algcount{937}},{\algcount{9212}},{\algcount{49061}}
\end{filecontents*}
\pgfplotstabletypeset[benchmarktable =
  {
    Algorithm & \multicolumn{6}{c}{\datasetname{ecoli}} \\
    & \multicolumn{3}{c}{\datasetname{acetate}} & \multicolumn{3}{c}{\datasetname{combined}}\\
    \cmidrule(lr){2-4} \cmidrule(lr){5-7}
    & {$c = 5$} & {$c = 7$} & {$c = 10$} & {$c = 5$} & {$c = 7$} & {$c = 10$} \\
  }]{data/cutoff.ecoli.csv}
  \algcaptioncutoff{ecoli}{networks \datasetname{acetate} and \datasetname{combined}}
  \label{tab:results-cutoff.ecoli}
 \end{sidewaystable}

\subsection{Discussion}
\label{s:discussion}
As shown in \cref{s:results}, the algorithms \algname{MMCS} and \algname{RS} from \cite{murakami2014} and \algname{HTC-BDD} from \cite{toda2013} are far faster than their competitors across a variety of input set families.

\algname{HTC-BDD} is extremely fast on inputs for which it terminates, outperforming its closest competitors by a factor of $4$ to $10$ on many inputs.
However, it frequently exhausted the 32GB available memory on our workstation.
In addition, it does not support cutoff enumeration.
Thus, we recommend \algname{HTC-BDD} for situations where all the MHSes of moderately-sized set families must be found quickly---for example, when many such families must be processed.
Since the core algorithm takes a ZDD representation of the input set family and returns either a BDD or a ZDD of its hitting sets, it is also very suitable for processing pipelines where BDDs are already used.

\algname{MMCS} and \algname{RS} are also very fast, and they support both cutoff\footnote{Supported by the first author's implementations \algname{pMMCS} and \algname{pRS}.} and full enumeration.
They have the additional benefit of very low memory requirements---in principle, the space required for a run depends only on the size of the input set family.
This is especially useful for inputs like \datasetname{HER2+.SHORT} where $S$ is small ($\approx$ 500 sets) but has an enormous collection of MHSes ($\approx$ 128 million).
Thus, we recommend these algorithms for situations where very large set families are studied or where only the small MHSes are required.

We note, however, that the provided implementations (both those by Murakami and Uno and by the first author) store the result MHSes in memory before writing them to disk, which did result in memory exhaustion for some inputs in our experiments.
It would be straightforward to modify the implementations of \algname{MMCS} or \algname{RS} to stream the result MHSes to disk rather than storing them in memory or to count them without storing them at all, as we did to compute the number of MHSes of size $\leq 10$ for \datasetname{HER2.short} and \datasetname{HER2.all} in \cref{tab:results-cutoff.ocsana}.
In addition, the implementations \algname{mmcs} and \algname{rs} of Uno and \algname{pMMCS} and \algname{pRS} of the first author varied dramatically in performance depending on the input, highlighting the importance of implementation.
Researchers planning to use any of these algorithms should certainly benchmark all the available implementations on data drawn from their application before adopting one.

We also find that parallel algorithms for MHS generation can be highly effective.
For example, the \algname{\MHStwo} algorithm (cf.~\cref{s:alg-mhs2}) of Cardoso and Abreu \cite{cardoso2014} shows a $2.33\times$ improvement in running time when passing from one thread to eight on the \datasetname{ecoli-acetate} set family, while the first author's parallel implementation \algname{pMMCS} of the \algname{MMCS} algorithm (cf.~\cref{s:alg-shd}) of Murakami and Uno \cite{murakami2014} shows a $4.04\times$ speedup when passing from one thread to eight on the \datasetname{ecoli-combined} set family.
Unfortunately, the first author's implementation of the \algname{BM} full-cover-based parallel algorithm (cf.~\cref{s:alg-pbm}) of Boros and Makino \cite{boros2009} was too slow to yield useful results.

\section{Conclusion}
\label{s:conclusion}
In this paper, we have surveyed the history and literature concerning the problem of generating minimal hitting sets.
The computational complexity of this task is a long-standing open problem.
However, since many applications (cf.~\cref{s:cognate}) depend on generating MHSes, a variety of algorithms (cf.~\cref{s:algorithms}) have been developed to solve it across numerous pure and applied research domains.

We have presented extensive benchmarks (cf.~\cref{s:results}) comparing the computation time required by nearly two dozen of these algorithms on a variety of inputs derived from real-world data.
These experiments consistently show that the \algname{MMCS} and \algname{RS} algorithms (cf.~\cref{s:alg-shd}) of Murakami and Uno \cite{murakami2014} and the \algname{HTC-BDD} algorithm (cf.~\cref{s:alg-htcbdd}) of Toda \cite{toda2013} are far faster than other available algorithms across a variety of inputs.
We have provided our benchmarking framework and code in easy-to-install Docker containers (cf.~\cref{s:methodology}), so researchers wishing to analyze the performance of these algorithms on their own inputs can do so easily.
Further details are available on our software repository at \url{https://github.com/VeraLiconaResearchGroup/MHSGenerationAlgorithms}.

\bibliographystyle{elsarticle-num}
\bibliography{sources}
\end{document}